\documentclass[review]{elsarticle}

\usepackage{lineno,hyperref}
\usepackage{color}
\usepackage{chemformula}
\usepackage{rotating}
\usepackage{graphicx}
\usepackage{colortbl}
\setchemformula{radical-radius=1pt}
\modulolinenumbers[5]

\journal{Icarus}

\biboptions{authoryear}

\begin{document}

\begin{frontmatter}

\title{Oxidation Processes Diversify the Metabolic Menu on Enceladus}


\author[UTSA,Swri]{Christine Ray \corref{mycorrespondingauthor}}
\cortext[mycorrespondingauthor]{Corresponding author}
\ead{christine.ray@contractor.swri.org}

\author[Swri]{Christopher R. Glein}

\author[Swri,UTSA]{J. Hunter Waite}

\author[Swri, UTSA]{Ben Teolis}

\address[UTSA]{Department of Physics and Astronomy, University of Texas, San Antonio, TX, USA}

\address[Swri]{Space Science and Engineering Division, Southwest Research Institute, San Antonio, TX, USA}

\author[ames]{Tori Hoehler}
\address[ames]{NASA Ames Research Center, Moffett Field, CA, USA}

\author[WHOI]{Julie A. Huber}
\address[WHOI]{Marine Chemistry and Geochemistry, Woods Hole Oceanographic Institution, Woods Hole, MA, USA}

\author[cornell]{Jonathan Lunine}
\address[cornell]{Department of Astronomy and Carl Sagan Institute, Cornell University, Ithaca, NY, USA}

\author[berlin]{Frank Postberg}
\address[berlin]{Institut f{\"u}r Geologische
Wissenschaften, Freie Universit{\"a}t Berlin, Berlin, Germany}

\begin{abstract}
The Cassini mission to the Saturn system discovered  a plume of ice grains and water vapor erupting from cracks on the icy surface of the satellite Enceladus. This moon has a global ocean in contact with a rocky core beneath its icy exterior, making it a promising location to search for evidence of extraterrestrial life in the solar system. The previous detection of molecular hydrogen (H$_2$) in the plume indicates that there is free energy available for methanogenesis, the metabolic reaction of H$_2$ with CO$_2$ to form methane and water. Additional metabolic pathways could also provide sources of energy in  Enceladus' ocean, but they require the use of other oxidants that have not been detected in the plume. Here, we perform chemical modeling to determine how the production of radiolytic O$_2$ and H$_2$O$_2$, and abiotic redox chemistry in the ocean and rocky core, contribute to chemical disequilibria that could support metabolic processes in Enceladus' ocean. We consider three possible cases for ocean redox chemistry: Case I in which reductants are not present in appreciable amounts and O$_2$ and H$_2$O$_2$ accumulate over time, and Cases II and III in which aqueous reductants or seafloor minerals, respectively, convert O$_2$ and H$_2$O$_2$ in the ocean to SO$_4^{2-}$ and ferric oxyhydroxides. We calculate the upper limits on the concentrations of oxidants and on the chemical energy available for metabolic reactions in all three cases, neglecting any additional abiotic reactions which could further affect energy availability. For all three cases, we find that many aerobic and anaerobic metabolic reactions used by microbes on Earth could meet the minimum free energy threshold, $\Delta G_{min}$, required for terrestrial life to convert ADP to ATP. We show that aerobic metabolisms could sustain up to $\sim 1$ cell cm$^{-3}$ within a 20 m depth across Enceladus' seafloor, even in our second case where O$_2$ and H$_2$O$_2$ are scarce. Additionally, anaerobic metabolisms could sustain up to $\sim 1$ cell for every two cm$^{-3}$ within this volume in our latter two cases. In contrast, methanogenesis could support up to $6 \times 10^{2}$ cells cm$^{-3}$ throughout this depth, due to the potential for a high hydrogen production rate at the seafloor as indicated by H$_2$ measurements from Enceladus' plume. While methanogenesis is the only metabolism that predicts cell density values close to  those reported in Earth's oceans and Antarctic subglacial lakes at this depth, our reported values depend on the area considered to be inhabited, which could be smaller than the entire Enceladus seafloor. Overall, the capacity for aerobic and anaerobic metabolisms to meet or exceed $\Delta G_{min}$ as well as sustain positive cell density values indicate that oxidant production and oxidation chemistry could contribute to supporting possible life and a metabolically diverse microbial community on Enceladus.

\end{abstract}


\end{frontmatter}


\section{Introduction}

Enceladus satisfies all the primary criteria for habitability, making it one of the most promising locations to search for extraterrestrial life in our solar system \citep{mckay2018enceladus}. Tidal heating created from the gravitational pull of Saturn and other Saturnian satellites maintains a liquid water ocean beneath its icy surface (\cite{iess2014gravity}, \cite{mckinnon2015effect}, \cite{thomas2016enceladus}, \cite{vcadek2016enceladus}). Constraints on the pH, temperature, and salinity of the ocean (\cite{glein2018geochemistry}, and references therein) are all within ranges tolerated by organisms on Earth. Compounds containing carbon, hydrogen, nitrogen, oxygen and possibly sulfur -- essential elements in life as we know it -- were all detected by Cassini's Ion Neutral Mass Spectrometer (INMS) and Cosmic Dust Analyzer (CDA) inside the plume over Enceladus' south polar region (\cite{waite2009liquid}, \cite{waite2017cassini}, \cite{postberg2018macromolecular}, \cite{postberg2018plume}). There is also chemical evidence that hydrothermal vents, another consequence of tidal heating \citep{choblet2017powering}, are present beneath the ocean, and could provide an additional source of chemical energy for life (\cite{hsu2015ongoing} and \cite{waite2017cassini}). These discoveries, which came from the Cassini mission, indicate that Enceladus is nominally, qualitatively habitable. The next step in understanding the biological potential of the Enceladus system is to evaluate the availability of resources quantitatively, to determine not just whether life might be able to survive there, but how abundant and productive it could be. These factors bear directly on the potential to detect life on Enceladus, as well as the strategies that could be called for to do so on a future mission there.

Energy availability has previously been discussed as a factor that could limit biology on ocean worlds (\cite{reynolds1983habitability}, \cite{gaidos1999life}, \cite{chyba2001life}, \cite{hand2007energy}, \cite{vance2016geophysical}). Compounds that could be used in metabolic reactions must be present in disequilibrium concentrations, such that biology can extract energy from the environment to drive the system toward equilibrium (\cite{mccollom1997geochemical}, \cite{shock2010potential}). After Cassini's E21 flyby through Enceladus' plume, \cite{waite2017cassini} reported that all of the compounds required for methanogenesis, a metabolic redox reaction which oxidizes molecular hydrogen (H$_2$) with carbon dioxide (CO$_2$) to form methane (CH$_4$) and water, were present in disequilibrium concentrations (chemical affinity of $\sim$50-120 kJ/mol, depending on modeled ocean pH), making methanogenesis a viable chemical energy source for life. While \cite{waite2017cassini} and other past studies (\cite{steel2017abiotic}, \cite{taubner2018biological}) have discussed the production of H$_2$ in the context of energy availability for methanogenesis on Enceladus, it is possible that other metabolic pathways could be viable energy sources for life there. Although CO$_2$ was the only oxidant detected by INMS, other oxidants may be present in the ocean at concentrations below the detection limits of Cassini's instruments, and these oxidants could create redox disequilibria in the ocean. Here, we examine how oxidant availability in Enceladus' ocean impacts the amount of chemical energy available there. 

When considering the relationship between biology and energy, two dimensions need to be evaluated. The chemical affinity, or the amount of free energy available from a metabolic reaction, must be sufficient to enable the energy to be captured and stored. In Earth's biology, this means the affinity must meet a threshold value, $\Delta G_{min}$, that is related to the energy required to phosphorylate ADP to ATP (\cite{hoehler2001apparent}, \cite{hoehler2004biological}). The second quantity that must  be considered is the flux of energy, which determines how much biomass can be supported in steady state (\cite{hoehler2004biological}, \cite{hoehler2013microbial}). To constrain the affinities and energy fluxes available from metabolic pathways other than methanogenesis, and therefore how oxidant production on Enceladus impacts biological potential there, the concentrations and fluxes of all redox species involved in these reactions must be considered. In this paper, we present a model of oxidant production within Enceladus to 1) constrain the oxidant budget, or the concentrations and fluxes of metabolically significant oxidants in the ocean, and 2) determine whether these additional metabolic pathways could provide sufficient energy for life. 


\section{Availability of Surface Oxidants}
Radiolysis as a mechanism for oxidant production on the surfaces of icy satellites has been discussed extensively (\cite{johnson2013sputtering}, \cite{teolis2017water} and references therein). These radiolytic oxidants form from the destruction of water molecules, and are then trapped at sites (e.g. defects, bubbles) within the water ice. In the south polar region of Enceladus where the surface is geologically active, radiolysis of surface ice may create oxidants which could be transported into the ocean. To evaluate this possible source of oxidants, we must first constrain the concentrations of O$_2$ and H$_2$O$_2$ in the ice. It is possible that O$_3$ may form in the ice as well, as it has been observed on other icy satellites in the solar system including the Saturnian satellites Dione and Rhea \citep{noll1997detection}, but it is formed by irradiation of O$_2$, and would thus likely be present in concentrations several orders of magnitude smaller than that of O$_2$ \citep{sittler2004pickup}. We will therefore only consider O$_2$ and H$_2$O$_2$ in this work. 

O$_{2}$ is produced most efficiently by non-penetrating, low-energy electrons ($<$400 eV) that break bonds mainly in the topmost surface layer ($\leq$3 nm) of the ice, from which radiolytic H$_2$ rapidly diffuses out allowing single oxygen atoms to recombine (\cite{teolis2005mechanisms}, \cite{teolis2009formation}). Across this small range, the energy flux in the ice, and the resulting concentration of O$_2$ that it produces, is approximately uniform. Trapped O$_2$ may be released as the ice surface is eroded (sputtered) by Saturn's plasma, unless the surface is buried by plume fallout substantially faster than it is sputtered away. Higher-energy electrons with greater penetration depths ($\geq$3 nm) deposit their energy deeper into the ice, where H and O can recombine, resulting mainly in reformation of water or radiolytic production of H$_{2}$O$_{2}$ at these depths (see \cite{teolis2017water}, Figure 10). 

Release of radiolytic O$_2$ by sputtering of the surface ice competes with deposition of fallback plume ice grains that bury the produced O$_2$. From the properties of the plasma environment at Enceladus, \cite{teolis2017water} calculated a surface sputtering rate of $3.0 \times 10^{12}$ H$_{2}$O molecules m$^{-2}$ $s^{-1}$, and a radiolytic O$_{2}$ sputtering rate of $7.5 \times 10^{11}$ O$_{2}$ molecules m$^{-2}$ $s^{-1}$ for a surface that is not buried, where all radiolytic products are ejected. Ice deposition rates range between 0.5 mm ice/year near the tiger stripe region ($\sim$30$^{\circ}$ of latitude from Enceladus' south pole, which corresponds to an area of $\sim 5.3 \times 10^4$ km$^{2}$) and 10 nm ice/yr at Enceladus' equator \citep{southworth2019surface}. If delivery occurs mostly over the tiger stripe region, where the surface is most active, and assuming an ice grain density of 0.9  g cm$^{-3}$, the reported global H$_2$O sputtering rate corresponds to a surface loss rate of 3 nm/yr in the tiger stripe region. This is smaller than all burial rates reported in \cite{southworth2019surface}, and much smaller than that in the tiger stripe region. Sputtering is therefore unable to remove a significant amount of O$_2$ produced in this region from the surface before it is buried. Rather than being sputtered, we assume that the radiolytic O$_2$ (i.e., that which is produced in the surface ice at a rate of $7.5 \times 10^{11}$ O$_{2}$ molecules m$^{-2}$ s$^{-1}$) can instead be delivered to the subsurface ocean. We assume that the O$_2$ concentration in the ice is constant with depth and increases with decreasing burial rate (dashed lines, Figure 1), with no probability for destruction as it moves through the ice shell. It is possible, however, that as O$_2$ travels downward through the ice, it may be altered and turned into H$_2$O$_2$ by reacting with H atoms also produced through radiolysis via the following pathway, described in \cite{teolis2017water}:

\begin{linenomath*}
\begin{equation}
H + O_{2} \rightarrow HO_2
\end{equation}
\begin{equation}
H + HO_2 \rightarrow H_2O_2 \\
\end{equation}
\end{linenomath*}

\noindent such that some of the amount of O$_2$ predicted may actually reach the ocean as H$_2$O$_2$. This uncertainty in the relative amounts of O$_2$ and H$_2$O$_2$ delivered to the ocean will be addressed in Section 4.

Unlike O$_2$, H$_2$O$_2$ is produced by penetrating electrons with energies greater than 400 eV and up to $6 \times 10^{3}$ keV, the largest particle energy detected by Cassini's Magnetospheric Imaging Instrument (MIMI) near Enceladus \citep{paranicas2012energetic}. This energy range corresponds to ice penetration depths of 3 nm (beyond which H$_2$ can no longer escape from the ice), up to 3 cm. We assume that only particles with vanishingly small energy flux (i.e. cosmic rays) penetrate beyond this depth, so that H$_2$O$_2$ molecules are no longer created or destroyed and the concentration is "frozen in". The energy flux of incident particles varies across this depth, resulting in production and destruction rates for H$_2$O$_2$ that also vary with depth. Additionally, burial of the surface by the plume acts to transport ice downward, meaning ice in geologically active regions is subject to radiation for a shorter period of time than a stationary surface. Thus, we must estimate the concentration of H$_2$O$_2$ as a function of its production and destruction rates at a given depth in the ice, $z$, as well as the surface burial rate. To estimate the H$_2$O$_2$ concentration in the ice, versus depth $z$ and for energies greater than 400eV, we take into consideration 1) the radiolysis yield $G_p$ (the number of H$_2$O$_2$ molecules produced per unit energy deposited), 2) the H$_2$O$_2$ destruction cross section $\sigma_D$, 3) the energy $E'$ of the electron with incident energy $E$ at depth $z$, 4) the electron stopping power $SP'$ at energy $E'$, and 5) the electron flux $F$ (per unit area, time and energy) at energy $E'$ that penetrates to depth $z$.  For the electron flux, $F$, we combine the  $\sim5$keV - 6 MeV electron distribution from the Cassini Magnetospheric Imaging Instrument (MIMI), reported in \cite{paranicas2012energetic}, with the  $<5$keV electron distribution from the Cassini Plasma Spectrometer (CAPS). For the latter we use a kappa distribution, which is commonly used to fit particle velocity distributions in space plasmas \citep{pierrard2010kappa}:

\begin{linenomath*}
\begin{equation}
f^{\kappa} = \frac{n_e}{(w_e\sqrt{\pi\kappa})^3} \frac{\Gamma(\kappa + 1)}{\Gamma(\kappa - \frac{1}{2})}\Big(1 + \frac{v_e^2}{\kappa w_e^2}\Big)^{-(\kappa + 1)}
\end{equation}
\end{linenomath*}

\noindent with an electron density, $n_e$, of $7.6 \times 10^7$ m$^{-3}$, estimated from the sum of the water group ion density calculated from \cite{wilson2008cassini} and the proton density from  \cite{sittler2006cassini}. We derive the electron thermal velocity, $w_e = (2T_e/m)^{\frac{1}{2}}$, from an electron temperature, $T_e$, of 10 eV \citep{sittler2006cassini}, and use $\kappa = 6$ \citep{schippers2008multi}. 

Electron stopping power, $SP'$, in eV/\AA~as a function of depth in ice can be found in \cite{laverne1985range} and \cite{ESTARdatabase}, and we approximate (neglecting particle straggling) the electron energy $E'$ at depth $z$ as $E - \int SP' \cdot dz$.  We first calculate the amount of hydrogen peroxide produced in a given volume element over time (molecules/cm$^3$/s), $P$, by:

\begin{linenomath*}
\begin{equation}
P = \int G_p \cdot F \cdot SP' \cdot dE. 
\end{equation}
\end{linenomath*}

\noindent Here we use $G_p = 6 \times 10^{-4}$ eV$^{-1}$ from \cite{teolis2017water}, appropriate for ice at a temperature of 80K (i.e. ice not in close proximity to the tiger stripes fissures).

The rate at which an H$_2$O$_2$ molecule is destroyed by incident radiation, $D$, in s$^{-1}$, is given by:

\begin{linenomath*}
\begin{equation}
D = \int \sigma_D \cdot F \cdot dE
\end{equation}
\end{linenomath*}

\noindent where we use $\sigma_D = 10^{-14.1}$ cm$^2$ from \cite{teolis2017water}. Without burial, the ratio of $P$ to $D$ yields the steady-state concentration of H$_2$O$_2$ molecules per unit volume at depth $z$. However in the the south polar region, where this concentration is diluted by burial from the plume, the rate at which the concentration of H$_2$O$_2$ changes in the ice is given by:

\begin{linenomath*}
\begin{equation}
\frac{dn_p}{dt} = -v\frac{dn_p}{dz} + P - D \cdot n_p,
\end{equation}
\end{linenomath*}

\noindent where $v$ is the rate of ice deposition from the plume in cm s$^{-1}$, and ${dn_p}/{dz}$ is the change in hydrogen peroxide concentration with depth. To etimate how much hydrogen peroxide reaches the ocean, we calculate the concentration of H$_2$O$_2$ down to 3$ \times 10^8$ \AA~(3 cm), which corresponds to a maximum particle energy of $6 \times 10^3$ keV \citep{paranicas2012energetic}. For the surface layer of ice to reach this level takes $\sim60$ years, given the current plume induced burial rate of 0.5mm/yr found at Enceladus' south pole \citep{southworth2019surface}. This solution to equation 6 at a given depth $z$', for which production, destruction and burial are balanced (${dn_p}/{dt} = 0$), is given by:

\begin{linenomath*}
\begin{equation}
n_p = \frac{1}{v}\int_{0}^{z'} (P - D\cdot n_p) dz.
\end{equation}
\end{linenomath*}

\begin{figure}[ht!]
\centerline{\includegraphics[height=2.5in]{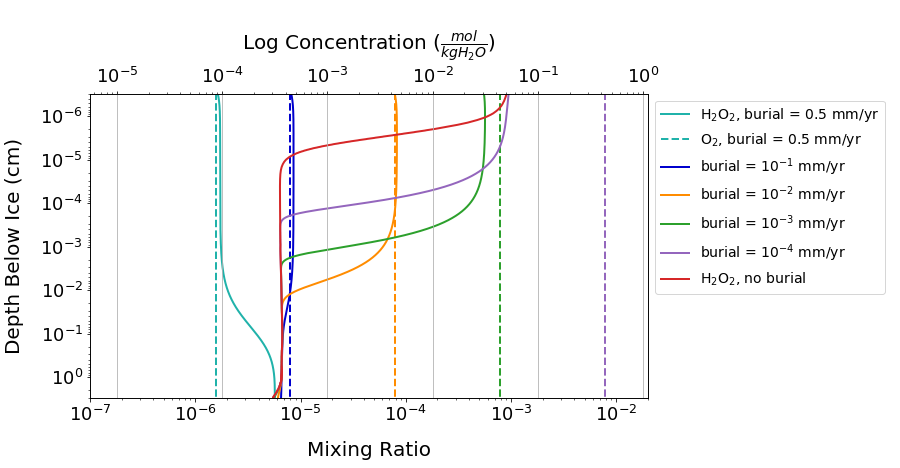}}
\caption{Mixing ratio and concentration of H$_2$O$_2$ (solid lines) and O$_2$ (dashed lines) with respect to H$_2$O from 3nm to 3cm below the ice surface for different plume burial rates. The steady-state H$_2$O$_2$ concentration for a surface with no plume burial is given in red for reference. Note that the steady-state O$_2$ concentration without burial is much higher, close to 20\% \citep{teolis2017enceladus}.}
\label{figone}
\end{figure}

The H$_2$O$_2$ concentration profile for various burial rates is shown in Figure 1, and is plotted with the O$_2$ concentration for comparison. Because the production and destruction of H$_2$O$_2$ occur over a much larger depth than the production and destruction of O$_2$, the resulting concentration profile of H$_2$O$_2$ in the ice varies with depth and this variation depends on the burial rate. Without burial, the H$_2$O$_2$ concentration is highest closest to the surface since the energy flux is also highest here, and drops off with depth until steady-state is reached (i.e., where production and destruction are balanced). With increasing burial rate, the uppermost concentration is diluted more and more, and steady-state is reached deeper in the ice shell since the ice is transported downward faster and faster. At the fastest burial rate reported in \cite{southworth2019surface}, ice still moves slowly enough that the H$_2$O$_2$ concentration reaches a value close to steady-state before H$_2$O$_2$ producing flux disappears (teal line, Figure 1).

\begin{figure}[ht!]
\centerline{\includegraphics[height=2.5in]{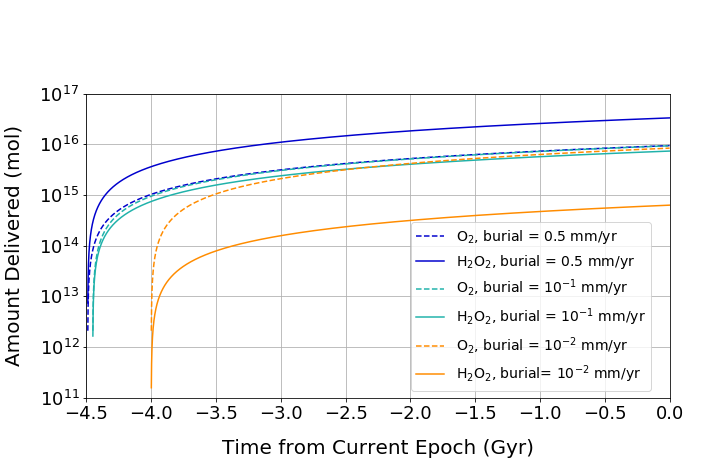}}
\caption{Cumulative delivery of O$_2$ (dashed lines) and H$_2$O$_2$ (solid lines) to Enceladus' ocean after 4.5 Gyr for different average plume deposition rates. Curve offsets reflect increased delivery time through the 5km thick ice shell, resulting from slower burial rates.} 
\label{figone}
\end{figure}

To determine how much O$_{2}$ and H$_{2}$O$_{2}$ could be transported into the ocean from a geologically active surface, we must turn the calculated concentration profiles into delivery rates. We consider ice deposition from the plume to be the primary surface vertical transport process at the tiger stripe region. The ice shell is also thinnest in this region ($\sim$5 km thick according to Cassini radar data reported in \cite{le2017thermally}). Assuming isostatic balance, the delivery period, or the time it takes for the uppermost layer of surface ice to initially reach the ocean, will thus be shortest in this region. For an average ice shell thickness of 5 km over this area, and assuming that the burial process is in steady-state (i.e. that the ice shell melts into the ocean as quickly as the surface ice is buried), this initial delivery period would be 10 Myr at the current plume fallout rate over the tiger stripe region (0.5 mm/yr). If the plume were less active over Enceladus geological history, only deposition rates greater than $\sim 2 \mu$m/yr could actually transport the surface ice to the ocean over the age of the solar system, which is the upper limit on Enceladus' age (\cite{mckinnon2018mysterious}, \cite{neveu2019evolution} and references therein).

Figure 2 shows how much O$_{2}$ and H$_2$O$_{2}$ would be delivered to the ocean from the tiger stripe region for a range of ice deposition rates over 4.5 Gyr, and through a 5 km thick ice shell. Lower deposition rates have greater curve offsets, which reflect increases in the initial delivery period. For the two slowest burial rates considered in Figure 1 ($10^{-3}$ and $10^{-4}$ mm/yr), the corresponding delivery periods exceed 4.5 Gyr, and thus these rates are not shown in Figure 2. In the case of O$_2$, increased delivery period is offset by increased concentration in the ice (Figure 1), such that the cumulative amount of O$_2$ delivered is similar for all burial rates, except for those too slow to deliver any oxidants within the 4.5 Gyr upper limit. Conversely, because the concentration of H$_2$O$_2$ is nearly the same for all burial rates (i.e. at or close to steady-state, Figure 1), the cumulative amount delivered decreases with slower burial. 

We take the upper limits on the total amounts of O$_2$ and H$_2$O$_{2}$ that could be delivered to the ocean from the ice shell, resulting from the highest surface turnover rate of 5 mm/yr, to be $9.5 \times 10^{15}$ and $3.4 \times 10^{16}$ mol, respectively. If the plume was not active or was less active and/or if the ice shell was thicker over the course of Enceladus' lifetime than at present day, then the delivery period required for the surface ice to reach the ocean would be greater than 10 Myr, and the total amount of oxidants delivered would be proportionally decreased. Conversely, if the plume were more active and/or the ice shell was thinner, more oxidant delivery could have occurred earlier on in Enceladus' history. Additionally, if the rate of vertical surface transport were smaller than the sputtering rate of surface ice, sputtering would remove O$_2$ and decrease the amount of net oxidants available. Finally, if Enceladus is younger than 4.5 Gyr, oxidant delivery would occur over a shorter period of time, decreasing the cumulative amounts predicted in Figure 2. The results of this calculation and subsequent calculations for a 100 Myr Enceladus, a lower limit on Enceladus' age based on estimates of the age of Saturn's rings \citep{cuzzi2018rings} and modeling of the orbital evolution of Saturn's inner satellites \citep{cuk2016dynamical}, are provided in Appendix A.


\section{Oxidant Production Internal to the Ocean}
Radiolysis of ocean water due to $^{40}$K decay is another source of oxidants on ocean worlds (\cite{chyba2001life}, \cite{altair2018microbial}). O$_{2}$, H$_{2}$ and H$_2$O$_{2}$ are created when water molecules are broken apart by high-energy particles that are emitted during the radioactive decay of $^{40}$K atoms dissolved in water. The detection of $\sim1.0 \times 10^{-3}$ mol K/kg H$_2$O in salty ice grains from the plume \citep{postberg2009sodium} suggests the presence of approximately $1.2 \times 10^{-7}$ moles of $^{40}$K/kg H$_2$O in the ocean, based on the present solar system abundance of $^{40}$K relative to total K \citep{lodders2003solar}, and assuming the concentration of K does not change between the ocean and the plume. With an ocean mass of approximately $2.8 \times 10^{19}$ kg, calculated from estimates of the ocean volume from \cite{vcadek2016enceladus}, this concentration is equivalent to $3.2 \times 10^{12}$ total moles of $^{40}$K. Past amounts of $^{40}$K in the ocean can be estimated using the radioactive decay law:

\begin{linenomath*}
\begin{equation}
n = n_{0} e^{-\lambda t},
\end{equation}
\end{linenomath*}

\noindent where $n$ represents the number of $^{40}$K moles at time $t$, $n_{0}$ is the number of moles at $t=0$ (i.e., when the solar system formed), and $\lambda = 5.54 \times 10^{-10}$ yr$^{-1}$ is the total radioactive decay constant for  $^{40}$K to both $^{40}$Ar and $^{40}$Ca. Solving for $n_{0}$ implies that there would have been $3.7 \times 10^{13}$ moles of  $^{40}$K if Enceladus existed and had a comparable ocean 4.5 Gyr ago.  
The production rates of radiolytic H$_{2}$, O$_{2}$ and H$_{2}$O$_{2}$ can be calculated at any point in time using the following rate equation:

\begin{linenomath*}
\begin{equation}
\frac{d[\textrm{X}]}{dt} = k_{X}[^{40}\textrm{K}]
\end{equation}
\end{linenomath*}

\noindent where brackets indicate the molal concentration of product species X, and $k_X$ corresponds to a first-order rate constant for the production of X. The concentration of $^{40}$K can be evaluated using Equation 8 if it is assumed that there has been no gain or loss of $^{40}$K from the ocean, and the mass of the ocean has been constant through time. The rate constants are determined here by parameterizing the production rates of H$_{2}$, O$_{2}$ and H$_{2}$O$_{2}$ from \cite{draganic1991decomposition}, who modeled the radiation chemistry of primitive ocean waters on Earth. The following values are derived: log$k_{H_2} = -5.87$,  log$k_{O_2} = -6.17$, and log$k_{H_2O_2}$ = -7.63 where all $k$ values are in units of yr$^{-1}$.

\begin{figure}[ht!]
\centerline{\includegraphics[height=3.0in]{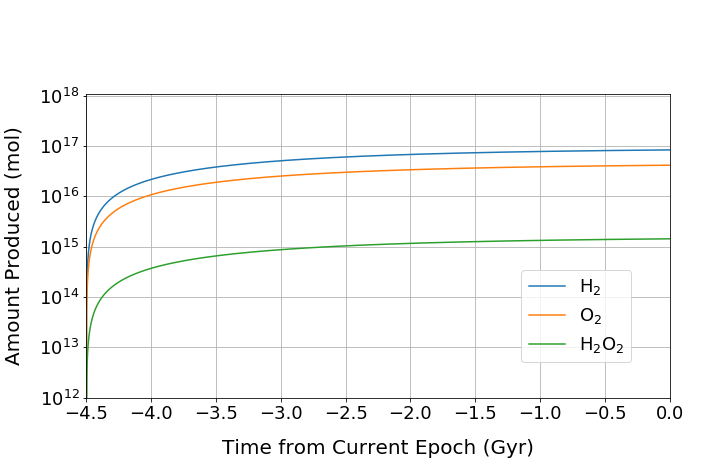}}
    \caption{Cumulative moles H$_2$ (blue), O$_2$ (orange), and H$_2$O$_2$ (green) produced through radiation chemistry by the decay of $^{40}$K in Enceladus' ocean.}
\label{figone}
\end{figure}
We can calculate the cumulative amount of each radiolytic product by combining Equations 8 and 9 and integrating:

\begin{linenomath*}
\begin{equation}
[\textrm{X}] = \frac{k_x[^{40}\textrm{K}]_0}{\lambda}(1-e^{-\lambda t})
\end{equation}
\end{linenomath*}

\noindent which can be evaluated as a function of time (Figure 3). Oxidant production slows with time as $^{40}$K is depleted. As an endmember case we again assume t = 4.5 Gyr. This provides upper limits on the amounts of O$_2$ ($4.2 \times 10^{16}$ mol) and H$_2$O$_2$ ($1.4 \times 10^{15}$ mol) from radiolysis in the ocean.

Here, we have assumed our rate constants do not differ significantly from those in \cite{draganic1991decomposition}, who consider an open system where any H$_2$ and O$_2$ produced can diffuse out of the ocean, making recombination of radiolytic products into H$_2$O$_2$ or H$_2$O less likely. In the case of Enceladus, where H$_2$ may be present in much higher concentrations than in Earth's primitive oceans, it is possible that the relative abundance of O$_2$ compared to H$_2$O$_2$ could be different, or that the oxidants produced could back-react with H$_2$ to form H$_2$O. For any O$_2$ and H$_2$O$_2$ that are delivered from the ice shell, laboratory studies of H$_2$-O$_2$-H$_2$O$_2$ equilibria \citep{foustoukos2011kinetics} indicate that back-reactions of oxidants with H$_2$ may be considerably slower at low temperatures than reduction by other species that are likely to be present in the ocean (i.e. ferrous iron and sulfides, discussed in the next section). These experiments also showed that elevated levels of H$_2$ at temperatures below 100$^{\circ}$C could slow the rate of H$_2$O$_2$ decomposition in water, making concentrations of H$_2$O$_2$ even higher than would be expected for a system without H$_2$. However, the H$_2$/O$_2$ ratio in Enceladus' ocean is likely much higher than that in the \cite{foustoukos2011kinetics} experiments. This could mean that H$_2$ will be a more powerful reductant on Enceladus than their kinetics would predict. Because there is a lack of kinetic studies of H$_2$-O$_2$-H$_2$O$_2$ equilibria in systems where H$_2$ is present in far higher concentrations than the oxidants, we have not incorporated reduction by H$_2$ into our model, but note that it could deplete the amount of oxidants that we have predicted to be delivered through the ice shell by reducing O$_2$ and/or H$_2$O$_2$ into H$_2$O. 

For oxidants produced by the decay of oceanic potassium-40, elevated concentrations of H$_2$ and other reductants compared to primitive Earth could be even more consequential. Intermediate species formed before O$_2$ and H$_2$O$_2$ (e.g. \ch{OH^.}, \ch{H^.}, hydrated electrons) are more reactive than their final products. Laboratory experiments which examined the effect of H$_2$ on radiolytic oxidant production in water found that small amounts of H$_2$ ($\sim10^{-5}$ M) quickly drove any O$_2$ and H$_2$O$_2$ produced to low steady-state concentrations ($10^{-12}$ and $10^{-8}$ M, respectively)  \citep{bjergbakke1989radiolytic}, further supporting the potential for elevated concentrations of H$_2$ to interfere with oxidant production on Enceladus. However, the situation may be more complicated because there are other reduced species in Enceladus ocean that could also compete for radiolytic intermediates, and their reactions would maintain net oxidant production. Observations from Cassini INMS indicate that methane is present on Enceladus in concentrations comparable to H$_2$ \citep{waite2017cassini}. Additional experiments reported in \cite{bjergbakke1989radiolytic} show that methane can also compete for oxidizing intermediates to form oxygenated organic species (some of which could be of potential prebiotic importance) at rates comparable to reactions of radiolytic products with H$_2$. Ferrous iron, sulfur-bearing species and organics, though present at smaller concentrations than H$_2$ and CH$_4$, can also compete for radiolytic intermediates (\cite{bjergbakke1989radiolytic}, \cite{bjergbakke1989radiolytic2}, \cite{draganic1991decomposition}). Further evidence that H$_2$ is not the only viable competitor for oxidizing intermediates comes from studies of subsurface water-rock systems. These have shown that sulfate production, likely driven by radiolysis, still occurs in systems with large amounts of aqueous H$_2$ \citep{li2016sulfur}. Such net oxidant production implies that back-reactions with H$_2$ are not completely efficient in geologic environments, presumably because the chemical complexity of these systems greatly exceeds what can be simulated in the laboratory. 

In summary, because the concentrations of reductants (namely H$_2$ and CH$_4$) in Enceladus' ocean are likely higher than those that were used in the \cite{draganic1991decomposition} model, a fraction of the reactive intermediates that create oxidants in our model will likely back-react with H$_2$ into H$_2$O, while others will be turned into O-bearing organics, sulfate, and ferric iron. It should be emphasized that the total amounts of O$_2$ and H$_2$O$_2$ that we have estimated to be delivered from the ice shell (Figure 2) and produced by $^{40}$K decay (Figure 3) are strictly upper limits, and oversimplify the network of reactions that will ultimately determine the fate of oxidants in Enceladus' ocean. We will further discuss how back-reactions and competition for oxidant-forming intermediates will impact our calculations in Section 4.1.


\section{The Ocean's Oxidant Budget}

Once O$_2$ and H$_2$O$_2$ are delivered to or produced in the ocean, they can either build up in the ocean or react abiotically with reductants. The rates at which aqueous reductants react with oxidants and/or are supplied to the ocean by seafloor minerals, and the extent of hydrothermal circulation through the rocky core (where oxidants can react with reduced minerals), determine which possibility is more realistic. If the resupply rates of aqueous reductants to the ocean are much slower than the production rate of oxidants, or if Enceladus is not hydrothermally active, then O$_2$ and H$_2$O$_2$ will build up in the ocean where they could potentially be used by life. If the resupply rates of aqueous reductants are sufficiently higher than the production rate of oxidants, or if oxidants can circulate through the rocky core, then the concentrations of O$_2$ and H$_2$O$_2$ in the ocean will be lower. We explore all of these possibilities through three cases, each of which encompass several different scenarios for oxidant production. A summary of all the oxidant budget cases and oxidant production scenarios explored in this section is provided in Table 1.

 \begin{table}[ht!]
\caption{A summary of the three oxidant budget cases which we explore in Sections 4.1, 4.2 and 4.3. Within each case, there are also several different scenarios for how oxidants might be produced, which will be introduced in Section 4.1}
\centering
    \hspace*{-1cm}
\begin{tabular}{|l|}
    \hline
    \textbf{Oxidant Budget Cases:} \\
    \hline
    \textbf{Case I:} Oxidants build up in the ocean; no reduction occurs \\
    \textbf{Case II:} Oxidants react with aqueous reductants in the ocean \\
    \textbf{Case III:} Oxidants react with reduced seafloor minerals \\
    \hline
    \textbf{Oxidant Production Scenarios:} \\
    \hline
    \textbf{Scenario 1:} Ice shell and ocean oxidant sources; only O$_2$ available\\
    \textbf{Scenario 2:} Ice shell and ocean oxidant sources; only H$_2$O$_2$ available \\
    \textbf{Scenario 3} No ocean oxidant source; only O$_2$ available (\textbf{3a}), or only H$_2$O$_2$ available (\textbf{3b})\\
    \hline
    \end{tabular}
\end{table}

\subsection{Case I: Buildup of O$_2$ and H$_2$O$_2$}

\begin{figure}[ht!]
\centerline{\includegraphics[height=2.8in]{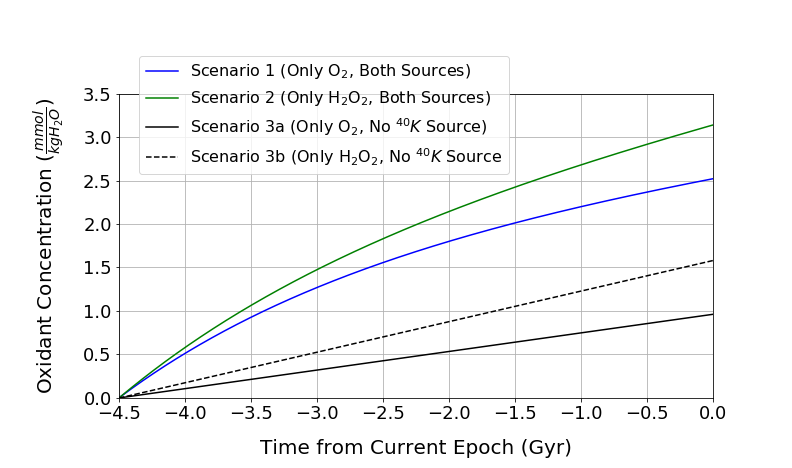}}
\caption{Concentration of O$_2$ or H$_2$O$_2$ in the ocean, assuming it has not been depleted by reactions with reductants. Here, we show three scenarios for oxidant production: 1) radiolysis (in both the ice shell and in the ocean) forms only O$_2$ (solid blue line), 2) radiolysis forms only H$_2$O$_2$ (solid green line), or 3) no oxidants are formed by $^{40}$K decay as a result of back-reactions with H$_2$ into H$_2$O so that the ice shell is the only source of oxidants to the ocean, and yields either all O$_2$ (3a, solid black line) or all H$_2$O$_2$ (3b, dashed black line). }
\label{figone}
\end{figure}

In our first end-member case, we assume that ocean water containing O$_2$ and H$_2$O$_2$ does not circulate through the rocky core, and that reductants in the ocean are either produced too slowly by water-rock interactions at the seafloor or react too slowly with oxidants at low temperatures to significantly affect the concentrations of these oxidants, allowing them to accumulate over time. As explained in Sections 2 and 3, there are several sources of uncertainty in the net amount of oxidants that could build up in this case, which must be addressed before the final concentrations of O$_2$ and H$_2$O$_2$ can be determined. In the ice shell, O$_2$ that is made at the surface could be further altered into H$_2$O$_2$ during transport to the ocean, which could shift the amounts of O$_2$ and H$_2$O$_2$ predicted in Section 2 (Figure 2). Additionally, although we have assumed in this case that reductants will not significantly affect the concentrations of O$_2$ and H$_2$O$_2$ after they have been produced, we cannot neglect the possibility that $^{40}$K decay in the ocean may not result in the production of these oxidants if H$_2$, CH$_4$ and other species detected by Cassini are present, as discussed in Section 3. Thus, to account for these uncertainties throughout the remainder of this work, we assume that any O$_2$ or H$_2$O$_2$ produced in the ice shell is not further processed by radiation after entering the ocean, and oxidant production on Enceladus results in three possible scenarios which are summarized in Table 1: 1) radiolysis (in both the ice shell and in the ocean) forms only O$_2$, 2) radiolysis forms only H$_2$O$_2$, or 3) no oxidants are formed by $^{40}$K decay as a result of back-reactions with H$_2$ into H$_2$O, making the ice shell the only source of oxidants to the ocean. In reality, as discussed in Section 3, the process of radiolysis by $^{40}$K decay would be far more complicated and could include many other end-products besides water, O$_2$ and H$_2$O$_2$, such as ferric iron, sulfate and oxidized organics, but the exact roles that these other species could play are difficult to determine without more definitive observational constraints on the types of organics that are present in the ocean, and a detailed kinetic model of ocean radiolysis that is beyond the scope of this work. 

In scenario 1, we estimate the amount of O$_2$ that could be present in the ocean by assuming that any H$_2$O$_2$ delivered from the ice shell dissociates into O$_2$ via 2H$_2$O$_2$ $\rightarrow$ O$_2$ + 2H$_2$O, such that the total amount of O$_2$ from the ice shell is equal to the amount of O$_2$ predicted in Figure 2 plus half the amount of H$_2$O$_2$. We also assume that radiolysis through $^{40}$K decay produces only O$_2$ and no H$_2$O$_2$, which is consistent with the prediction shown in Figure 3, where H$_2$O$_2$ comprises only 3\% of all oxidants produced. In scenario 2, we estimate the amount of H$_2$O$_2$ in the ocean by assuming that any O$_2$ made in the ice shell was converted to H$_2$O$_2$ as it was transported through the ice via the pathway in Equations 1 and 2. Since this is a one-to-one conversion between O$_2$ and H$_2$O$_2$, the total amount of H$_2$O$_2$ delivered from the ice shell is then equal to the amount of H$_2$O$_2$ plus the amount of O$_2$ predicted in Figure 2. We further assume that radiolysis resulting from $^{40}$K decay produces mostly H$_2$O$_2$ (and negligible amounts of O$_2$), which could be another possible result of elevated amounts of H$_2$ in the ocean. In scenario 3, we assume that all radiolytic oxidizing intermediates have reacted with hydrothermally sourced H$_2$ to form water, such that there is no production of either O$_2$ or H$_2$O$_2$ from $^{40}$K decay, and that the only oxidants available are from the ice shell. In this case, we again assume that either all O$_2$ has been made (line 3a, in Figure 4), or all H$_2$O$_2$ has been made (line 3b, in Figure 4), with these amounts calculated in the same way as in scenarios 1 and 2.

For each end-member scenario, we calculate the theoretical concentration of oxidants in the ocean over time based on the rates of production at the south polar surface and by $^{40}$K decay in the ocean. If Enceladus resembles Case I, we assume the most realistic concentration of each of these oxidants falls below the upper limits predicted in each scenario (i.e., both O$_2$ and H$_2$O$_2$ are present in the real Enceladus ocean). The cumulative concentrations of oxidants in the ocean, present as either O$_2$ or H$_2$O$_2$, are shown in Figure 4, and could reach as high as 2.5 mmol O$_2$/kg H$_2$O or 3.1 mmol H$_2$O$_2$/kg H$_2$O with both an ice shell and $^{40}$K source (Scenarios 1 and 2), and 1.0 mmol O$_2$/kg H$_2$O or 1.6 mmol O$_2$/kg H$_2$O if there is only an ice shell source (Scenarios 3a and 3b).


\subsection{Case II: Oxidation of Aqueous Reductants}

In our second and third end-member cases, we consider an environment in which reductants are available in abundance to react with any oxidants produced in or delivered to the ocean (Table 1). If the chemical composition of Enceladus' core contains chondritic abundances of rock-forming elements (\cite{zolotov2007oceanic}, \cite{sekine2015high}), the two most abundant elements that could be present in reduced forms should be sulfur and iron. Laboratory studies of O$_2$ and H$_2$O$_2$ reduction by ferrous iron and sulfide have yielded kinetic equations for these reactions that can be extrapolated to Enceladus ocean conditions, showing that these reductants could affect the concentrations of these oxidants, even at 0 $^{\circ}$C (see section 4.2.2). As mentioned in Section 3, Cassini observations indicate that H$_2$ and CH$_4$ \citep{waite2017cassini}, as well as hydrogen-bearing organics (\cite{khawaja2019low}, \cite{postberg2018macromolecular}, \cite{magee2017neutral}) are also likely present in the ocean. For H$_2$, results from laboratory experiments reported in \cite{foustoukos2011kinetics} indicate that its ability to reduce O$_2$ and H$_2$O$_2$ may be kinetically inhibited at low (T = 0$^{\circ}$C) temperatures, although their results cannot be directly extrapolated to the relatively high H$_2$/O$_2$ and H$_2$/H$_2$O$_2$ ratios we predict in this work. Because there do not appear to be any other kinetic models available that can be directly extrapolated to Enceladus ocean conditions, we will continue to assume that reactions with H$_2$ can be neglected, although they could play a more important role in the real Enceladus system. For CH$_4$, we again do not have kinetic information available that can be directly applied to the conditions in our model, but experiments indicate that its oxidation by both O$_2$ (\cite{webley1991fundamental} \cite{shilov2001activation} \cite{sorokin2010oxidation}) and H$_2$O$_2$ (\cite{seki2000reaction} \cite{park2003copper} \cite{ab2013oxidation}) is kinetically inhibited at low temperatures, so we will assume that this will also be the case on Enceladus. 

In the case of organics, their influence on the oxidant budget is even more uncertain.  Highly refractory organics are encapsulated in ice grains that form just above the ocean level, and are transported out of surface vents and detected by the CDA instrument as high mass organic cations (HMOC), which have a complex molecular structure that cannot be quantitatively related to their ocean concentration \citep{postberg2018macromolecular}. On the other hand, for volatile species detected in the plume by the INMS instrument \citep{magee2017neutral} and CDA \citep{khawaja2019low}, transport through cold ice in the plume is a significant process for fractionation of the organic composition by adsorption \citep{bouquet2019adsorption}. Another complication is fragmentation of molecules in the INMS antechamber during high-speed sampling \citep{waite2009liquid}. All of these factors make it difficult to constrain exactly which types of organics are present in Enceladus' ocean and in what concentrations. As with H$_2$, there is also some uncertainty in how organic redox reactions would compare kinetically to those for ferrous iron and sulfide in our system. For O$_2$, \cite{chang1999coal} found that the oxidation rate of sedimentary organic matter is two to three orders of magnitude smaller than pyrite oxidation in air. While these findings are again for an environment different from that of the Enceladus ocean, they do provide some justification in choosing to exclude this reaction from our model. For H$_2$O$_2$, however, reactions with organic matter could be more important. There have been numerous studies that have found H$_2$O$_2$ to be an effective oxidant of various organic compounds (\cite{satterfield1954reaction}, \cite{mcdonald1998oxidation}), including at cold ($0^{\circ}$C) temperatures (\cite{mikutta2005organic}, \cite{davila2008subsurface}). While some of these studies have provided kinetic data indicating that these reactions could be as rapid as those between H$_2$O$_2$ and ferrous iron or sulfide species at $0^{\circ}$C (\cite{mcdonald1998oxidation}, \cite{pasek2020plume}), as with H$_2$, they do not provide enough information to extrapolate those kinetics to our system. Without kinetic studies under conditions more analogous to the Enceladus ocean, and better constraints on both the types of organics present in the ocean and their concentrations, it would be premature to attempt to determine exactly how organics on Enceladus may affect the oxidant budget. We will therefore focus on other oxidation reactions that are well-characterized as a first step in our model, but we emphasize once again that the net oxidant concentrations that are calculated here are strictly upper limits. In the actual Enceladus system, any oxidants produced will likely be consumed not only by iron and sulfide species, but also by H$_2$ and organics, with organics likely being especially important for the H$_2$O$_2$ budget.

With ferrous and sulfide having been established as the key reductants that 1) are likely to react quickly enough with O$_2$ and H$_2$O$_2$ at freezing temperatures to affect their concentrations, 2) have kinetic data available in the literature that can be directly extrapolated to Enceladus ocean conditions, and 3) have chemical forms and concentrations in the ocean that can be constrained, we proceed in quantifying their effects on the oxidant budget. In our second end-member case (Case II), we assume that oxidants only react with aqueous sulfur and iron reductants that are dissolved in the ocean.  In our third end member case (Case III), we assume that ocean water can percolate into the rocky core, and that oxidants react directly with reduced iron and sulfur bearing minerals within the rocky core or at the seafloor.

\subsubsection{Constraining Reductant Abundances from Mineral Dissolution}
In Case II, we assume that reductants are present as total dissolved sulfide ($\Sigma$H$_2$S = H$_2$S + HS$^-$) and dissolved ferrous iron (Fe$^{2+}$) in the ocean, and that their resupply rates are sufficiently fast to maintain equilibrium concentrations of these reductants, such that their production is much faster than the production and delivery of oxidants. To constrain the amount of sulfide that could be in the ocean, we apply the upper limit on plume H$_2$S reported in \cite{magee2017neutral} of $<$100 ppm by volume, which corresponds to a molar mixing ratio of 0.01$\%$. If we scale this mixing ratio to CO$_2$, which has been used as a reference species for dissolved gases in the ocean due to its low condensibility in the tiger stripes (\cite{waite2017cassini}, supplemental material), this implies an upper limit on the ocean concentration of H$_2$S of $2 \times 10^{-6}$ mol/kg H$_2$O at pH 9, and $1 \times 10^{-9}$ mol/kg H$_2$O at pH 11. At these alkaline pH values, it is likely that HS$^{-}$ will be the more predominant form of dissolved sulfide and $\Sigma$H$_2$S$ \approx$ [HS$^-$]. While the CDA instrument on Cassini had the capability to detect HS$^-$, no observational upper limit is available on this species because the corresponding analysis has not taken place. However, if sulfide species are in equilibrium in the ocean, our estimated concentration of H$_2$S implies a HS$^-$ concentration of $9 \times 10^{-5}$ or $5 \times 10^{-6}$ mol/kg H$_2$O at pH 9 or pH 11, respectively.

 \begin{table}[h]
    \small
\caption{Concentrations of neutral and ion species detected by the INMS and CDA instruments, respectively, and reported in \cite{waite2017cassini}, \cite{postberg2009sodium}, \cite{postberg2011salt} and \cite{glein2018geochemistry}. All concentrations are in mmoles per kilogram of water.}
    \begin{tabular}{l c c}
	    \hline
	   Quantity & pH 9 & pH 11 \\
	    \hline
	    CO$_2$ & $7 \times 10^{-2}$ & $1 \times 10^{-4}$ \\
	    Na$^+$ & 130 & 154 \\
	    K$^+$ & 1.3 & 1.54\\
        \hline
	\end{tabular}
\end{table}

Because we do not have an observational constraint on reduced iron, we constrain its equilibrium concentration in the ocean by first determining which iron-bearing mineral could be controlling the production of this species. Given the constraints, reported here in Table 2, on the activity (or effective concentration) of CO$_2$ in the ocean reported in \cite{waite2017cassini}, and the plume ice grain concentrations of NaCl and KCl constrained in \cite{postberg2009sodium} and \cite{postberg2011salt} and adapted into the ocean model from \cite{glein2020carbonate}, we use Geochemist's Workbench and the default thermo database \citep{bethke2007geochemical} to calculate the solubilities of seafloor minerals at the expected ocean temperature (0$^{\circ}$C) and seafloor pressure (73 bar). We adopt a pH range of 9 to 11 (\cite{glein2018geochemistry}, \cite{waite2017cassini}), which encompasses the results of several previous modeling efforts (\cite{zolotov2007oceanic}, \cite{postberg2009sodium}, \cite{hsu2015ongoing}, \cite{glein2015ph}). We treat the activities of H$_2$ and silica (SiO$_2$) as free parameters to allow a generalized visualization of the phase relations. We apply the H$_2$ activity constraints from \cite{waite2017cassini} (dashed blue lines in Figures 5a and 5b) to guide the decision of which minerals to adopt as the buffering source of Fe$^{2+}$. For a pH = 9 ocean, an H$_2$ activity of $1 \times 10^{-4}$ indicates that, with increasing SiO$_2$ activity, either siderite or minnesotaite is predicted to be the most stable iron-bearing mineral in the system, and could act as a source of Fe$^{2+}$. At pH = 11, an H$_2$ activity of $2 \times 10^{-7}$ indicates that, with increasing $a$SiO$_2$, either magnetite, cronstedtite, greenalite or minnesotaite is predicted to be stable.

\begin{figure}[ht!]
\centerline{\includegraphics[height=3.0in]{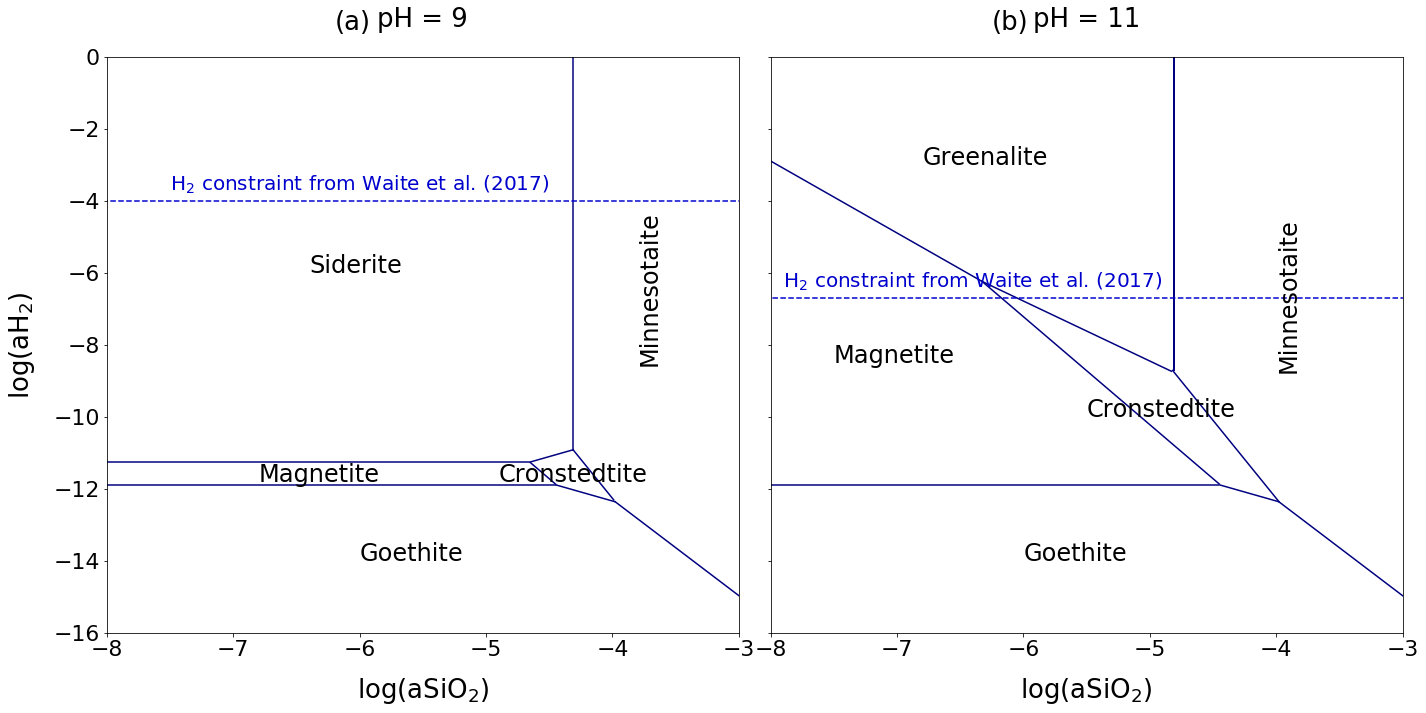}}
\caption{Stable iron-bearing minerals at 0$^{\circ}$C and 73 bar, as a function of the activity of SiO$_2$ at pH = 9 (a) and pH = 11 (b). In both plots we have applied the $a$CO$_2$ constraints at pH 9 or 11 from \cite{waite2017cassini}. Here we also show the $a$H$_2$ constraint from \cite{waite2017cassini} (dashed horizontal lines), which we use to guide the choice of iron-bearing minerals to adopt as the aqueous Fe$^{2+}$ buffers in our model. }
\label{figone}
\end{figure}

\begin{figure}[ht!]
\centerline{\includegraphics[height=2.5in]{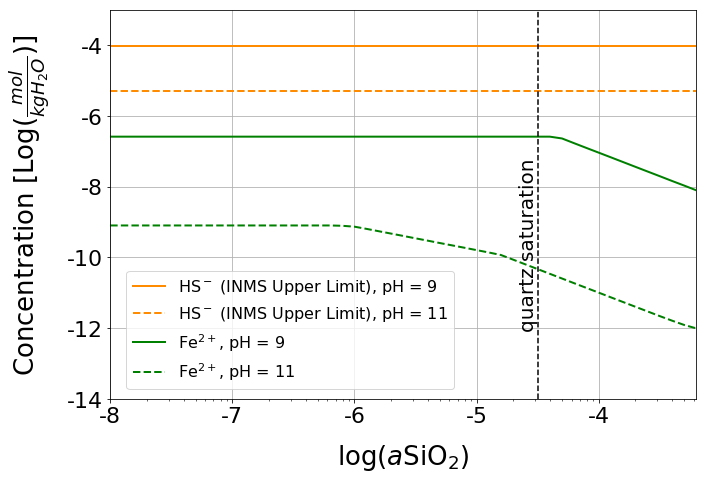}}
\caption{Equilibrium concentrations of dissolved ferrous iron as a function of silica activity, and HS$^-$ concentration derived from INMS observations, in Enceladus' ocean at pH 9 and pH 11. The concentrations of ferrous iron are determined by the dissolution of the stable minerals shown in Figure 5.} 
\label{figone}
\end{figure}

Assuming the presence of minerals that control the concentration of Fe$^{2+}$ (at fixed pH) in the ocean, we perform speciation calculations with Geochemist's Workbench \citep{bethke2007geochemical} to determine a range of values for [Fe$^{2+}$]. We apply the same constraints on aH$_{2}$S, $a$CO$_2$, [K$^+$], and [Na$^{+}$] described above, while allowing $a$SiO$_2$ to be a free parameter. The concentrations of Fe$^{2+}$ at pH 9 and pH 11 are shown in Figure 6. We end the calculation at $a$SiO$_2 = 10^{-3.2}$ because higher values lead to high concentrations of silicate anions at pH 11, but data from \cite{postberg2009sodium} indicates that chloride rather than silicate should be the dominant anion in the ocean.

\subsubsection{The kinetics of aqueous oxidation reactions}
With the ranges of concentrations for HS$^-$ and Fe$^{2+}$ constrained, we determine the partitioning of reactivity (i.e., branching ratios) for O$_2$ and H$_2$O$_2$ between each reductant via:

\begin{linenomath*}
\begin{equation}
4\textrm{Fe}^{2+} + \textrm{O}_2 + 6\textrm{H}_2\textrm{O} \rightarrow 4\textrm{FeOOH} + 8\textrm{H}^+,
\end{equation}
\begin{equation}
\textrm{HS}^- + 2\textrm{O}_2 \rightarrow \textrm{SO}_{4}^{2-} + \textrm{H}^+,
\end{equation}
\begin{equation}
2\textrm{Fe}^{2+} + \textrm{H}_2\textrm{O}_2 + 2\textrm{H}_2\textrm{O} \rightarrow 2\textrm{FeOOH} + 4\textrm{H}^+,
\end{equation} 
\begin{equation}
\textrm{HS}^- + 4\textrm{H}_2\textrm{O}_2 \rightarrow \textrm{SO}_{4}^{2-} + 4\textrm{H}_2\textrm{O} + \textrm{H}^+
\end{equation}
\end{linenomath*}

\noindent where we have taken FeOOH to be representative of ferric oxyhydroxide minerals. We use the rate laws given in \cite{millero1987oxidationFe}, \cite{millero1987oxidationH2S}, \cite{millero1989oxidationFe}, and \cite{millero1989oxidationHS}:
\begin{linenomath*}
\begin{equation}
-\frac{d[\textrm{O}_2]}{dt} = \frac{1}{4}k_{11}[\textrm{Fe}^{2+}][\textrm{OH}^-]^2[\textrm{O}_2], 
\end{equation}
\begin{equation}
-\frac{d[\textrm{O}_2]}{dt} = 2k_{12}[\textrm{H}_2\textrm{S}]_T[\textrm{O}_2], 
\end{equation}
\begin{equation}
-\frac{d[\textrm{H}_2\textrm{O}_2]}{dt} = \frac{1}{2}k_{13}[\textrm{Fe}^{2+}][\textrm{OH}^-][\textrm{H}_2\textrm{O}_2], 
\end{equation}
\begin{equation}
-\frac{d[\textrm{H}_2\textrm{O}_2]}{dt} = 4k_{14}[\textrm{H}_2\textrm{S}]_T[\textrm{H}_2\textrm{O}_2] 
\end{equation}
\end{linenomath*} 

\noindent where we have established that [H$_2$S]$_T \approx$ [HS$^-$] in Section 4.2.1, and the rate constants  $k_{11}$, $k_{12}$, $k_{13}$ and $k_{14}$, corresponding to equations 11-14, can be calculated by:
\begin{linenomath*}
\begin{equation}
 \textrm{log} k_{11} = 21.56 - 1545/T - 3.29I^\frac{1}{2} + 1.52I \textrm{ (M}^{-3} \textrm{min}^{-1}),
 \end{equation}
\begin{equation}
\textrm{log} k_{12} = 11.78 - 3000/T + 0.44I^\frac{1}{2} \textrm{ (M}^{-1} \textrm{h}^{-1}),
\end{equation}
\begin{equation}
\textrm{log} k_{13} = 11.72 - 2.4I^{\frac{1}{2}} + 1.38I \textrm{ (M}^{-2} \textrm{s}^{-1}),
\end{equation}
\begin{equation}
\textrm{log} k_{14} = 8.60 - (2052/T) - 0.084I^\frac{1}{2} \textrm{ (M}^{-1} \textrm{min}^{-1}).
\end{equation}
\end{linenomath*}

\noindent In the above equations, we take temperature $T$ to be 273 K, and the ionic strength $I$ to be $\sim$0.1 molal as constrained in \cite{glein2018geochemistry}. It should be noted that the rate equations for oxidation of sulfide both by O$_2$ (equation 16) and H$_2$O$_2$ (equation 18) do not explicitly include pH dependence. For oxidation by O$_2$, the rate becomes independent of pH above pH 8, according to \cite{millero1987oxidationH2S}. For oxidation by H$_2$O$_2$, we scale our value for log$k_{11}$ down by a factor of 0.9 at pH 9 and 0.7 at pH 11 to account for the decrease in rate with higher pH above pH = 8 as discussed in \cite{millero1989oxidationHS}.

The complete differential equations that describe how the concentrations of O$_2$ and H$_2$O$_2$ change in Enceladus' ocean over time, assuming that all oxidants are present as either O$_2$ or H$_2$O$_2$, can be constructed from equations 15-18, along with the two production terms describing transport from the ice shell and radiolysis of ocean water from $^{40}$K decay:

\begin{linenomath*}
\begin{equation}
\frac{d[O_2]}{dt} = P_{O_2,ice} + \frac{1}{2}P_{H_2O_2,ice} +  k_{40K}[^{40}K] - \big(\frac{1}{4}k_{11}[Fe^{2+}][OH^{-}]^2 + 2k_{12}[HS^-]\big)[O_2],
\end{equation}
\begin{equation}
\frac{d[H_2O_2]}{dt} = P_{O_2,ice} + P_{H_2O_2,ice} + k_{40K}[^{40}K] - \big(\frac{1}{2}k_{13}[Fe^{2+}][OH^-] +  4k_{14}[HS^-]\big)[H_2O_2].
\end{equation}
\end{linenomath*} 

\noindent Here, $P_{O_2, ice}$ and $P_{H_2O_2, ice}$ are the rates of O$_2$ and H$_2$O$_2$ delivery from the ice shell, respectively (equal to zero during the initial 10 Myr delivery period and $2.6 \times 10^6$ mol/yr and $7.4 \times 10^6$ mol/yr, respectively, thereafter). As described in Section 4.1 (Scenario 1), we have included a factor of 1/2 in front of the  $P_{H_2O_2, ice}$ term in Equation 23 because we have assumed that all H$_2$O$_2$ has dissociated into O$_2$ in the ocean via 2H$_2$O$_2$ $\rightarrow$ O$_2$ + 2H$_2$O . In Equation 24, we have assumed that any O$_2$ produced in the topmost surface of the ice has been converted to H$_2$O$_2$ by Reactions 1 and 2 in a one-to-one conversion (Section 4.1, Scenario 2).  In the third term on the right hand side of Equations 23-24, we have defined $k_{40K}$ = $k_{O_2} + k_{H_2O_2}$ from Equation 9 (Section 3), and $[^{40}K] = [^{40}K]_0 e^{-\lambda t}$. While we have not included an explicit equation for Scenario 3 (Section 4.1), where there is no oxidant source from $^{40}$K decay, we note that this third production term could be smaller than we predict here, if elevated concentrations of reductants such as H$_2$ inhibit oxidants from being formed via this pathway. The reductant loss terms at the ends of Equations 23-24 have been compiled from Equations 15-18, where we have assumed [H$_2$S]$_T \approx$ [HS$^-$].

\begin{figure}[ht!]
\centerline{\includegraphics[height=2.7in]{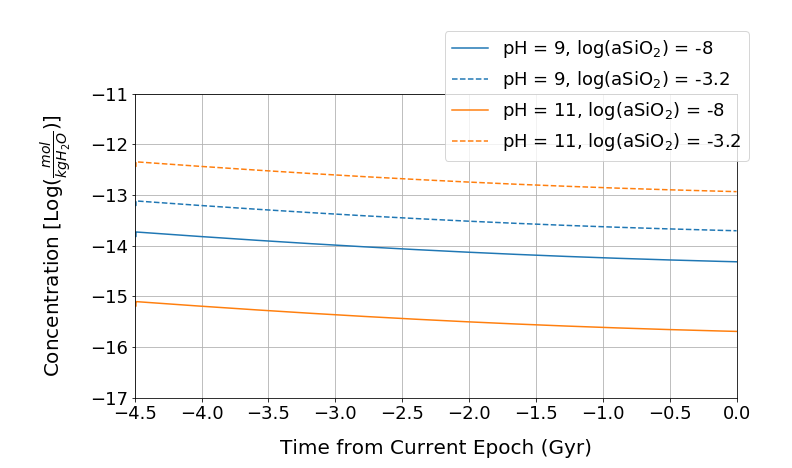}}
\caption{Steady-state concentration of O$_2$ over time. pH 9 concentrations are shown in blue, and pH 11 concentrations are shown in orange. Solid lines correspond to log(aSiO$_2$ = -8), and dashed lines to log(aSiO$_2$ = -3.2).}
\label{figone}
\end{figure}

\begin{figure}[ht!]
\centerline{\includegraphics[height=2.7in]{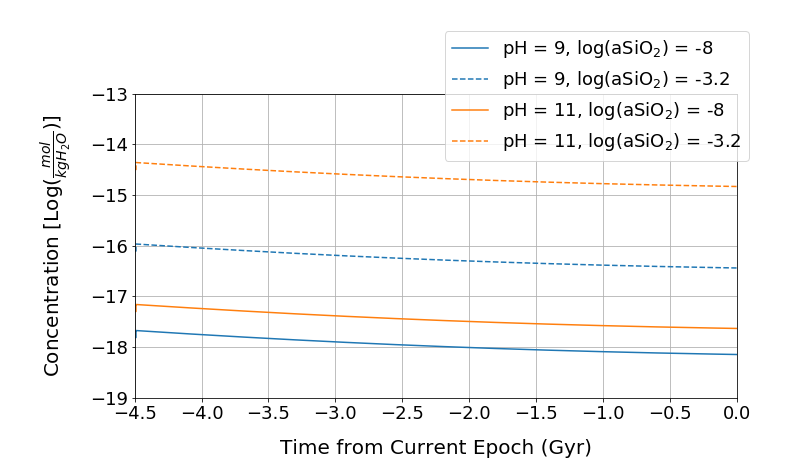}}
\caption{Steady-state concentration of H$_2$O$_2$ over time. pH 9 concentrations are shown in blue, and pH 11 concentrations are shown in orange. Solid lines correspond to log(aSiO$_2$ = -8), and dashed lines to log(aSiO$_2$ = -3.2).} 

\label{figone}
\end{figure}

Solving equations 23 and 24 numerically over very short time intervals ($\sim$1 year), both from $t=0$ and $t = 20$ Myr to account for the initial delivery period when both $P_{ice}$ terms equal 0, shows that the O$_2$ and H$_2$O$_2$ concentrations in the ocean are immediately driven to steady-state (the concentrations when the differential terms are set to zero), due to the loss terms dominating the system in both cases. These steady-state concentrations of O$_2$ and H$_2$O$_2$ over time are shown in Figures 7 and 8,  and are between $ 10^{-16}$ and $10^{-12}$ mol/kg H$_2$O and between $ 10^{-19}$ and $10^{-14}$ mol/kg H$_2$O, respectively. Equivalent figures for Scenario 3, where the ice shell is the only source of oxidants, are provided in Appendix B. In all cases, the resulting steady-state concentration of H$_2$O$_2$ is lower than the concentration of O$_2$ (i.e., H$_2$O$_2$ reacts faster) due to larger loss terms in equations 23 and 24.
 
As would be expected for an ocean where reductants are far more abundant than oxidants, the concentrations of O$_2$ and H$_2$O$_2$ are very low. However their abiotic consumption produces ferric oxyhydroxides and SO$_4^{2-}$, which could also be biologically useful oxidants. In the case where all oxidants are in the form of O$_2$, we partition the O$_2$ flux into consumption by Fe$^{2+}$ vs. HS$^{-}$ according to the rates of the respective processes (Equations 11-12 and 15-16), using the calculated steady-state concentration of O$_2$ through time:

\begin{linenomath*}
\begin{equation} 
 \frac{d[FeOOH]}{dt} = k_{11}[Fe^{2+}][OH^-]^2[O_2],
 \end{equation} 
\begin{equation}
\frac{d[SO_4^{2-}]}{dt} = k_{12}[HS^-][O_2]
 \end{equation}
\end{linenomath*} 

\noindent where we have again chosen FeOOH to be representative of ferric oxyhydroxide minerals. In the case where all oxidants are in the form of H$_2$O$_2$, we use (derived from equations 13-14 and 17-18):

\begin{linenomath*}
\begin{equation} 
 \frac{d[FeOOH]}{dt} = k_{13}[Fe^{2+}][OH^-][H_2O_2],
 \end{equation} 
\begin{equation}
\frac{d[SO_4^{2-}]}{dt} = k_{14}[HS^-][H_2O_2]. 
 \end{equation}
\end{linenomath*} 

\noindent The results for oxidation by O$_2$ and oxidation by H$_2$O$_2$ are plotted in Figures 9a-b and Figures 9c-d, respectively, with equivalent figures for Scenario 3 provided in Appendix B. We have assumed that FeOOH and SO$_4^{2-}$ accumulate in the ocean or at the seafloor as secondary oxidants, where they could be available for use by life, and are not abiotically reduced back to ferrous iron and sulfides. This may be a valid assumption for SO$_4^{2-}$, as available experimental data indicate that sulfate reduction is kinetically inhibited at more alkaline ($>6$) pH values (\cite{truche2009experimental} \cite{tan2019}). For ferric iron, the potential for abiotic reduction to occur under Enceladus ocean conditions could be greater. The ability for life to use ferric iron metabolically would depend on the relative rates of biotic versus abiotic iron reduction, as it does in systems on Earth (e.g., \cite{mortimer2011experimental}). Laboratory studies of microorganisms in sediments indicate that iron reduction may not be an important metabolic process for organisms living in alkaline environments \citep{marquart2019influence}, meaning abiotic iron reduction may be faster, although competition could still occur to some degree. This underscores the need for future experiments to constrain the abiotic rates of ferric iron reduction in the presence of potential electron donors (H$_2,$ FeS, organic matter) under Enceladus ocean conditions \citep{glein2018geochemistry}.

Although the reductant loss terms for H$_2$O$_2$ in equation 24 are larger than those for O$_2$ in equation 23, it is evident from Figure 9 that more of the oxidized products form from O$_2$ than from H$_2$O$_2$ due to reaction stoichiometry (one mole of O$_2$  produces twice as much FeOOH or SO$_4^{2-}$ as one mole of H$_2$O$_2$, Equations 11-14). However, in both cases the maximum abundance of FeOOH ($\sim$10 mmol/kg H$_2$O) corresponds to a situation in which almost all of the O$_2$ or H$_2$O$_2$ available is converted to FeOOH. The only set of considered conditions in which ferrous iron oxidation is not strongly favored, and the amount of SO$_4^{2-}$ produced is comparable to the amount of FeOOH produced, is oxidation by O$_2$ at pH = 9 and log(aSiO$_2$) = -3.2. 

\begin{figure}[ht!]:
\centerline{\includegraphics[height=4in]{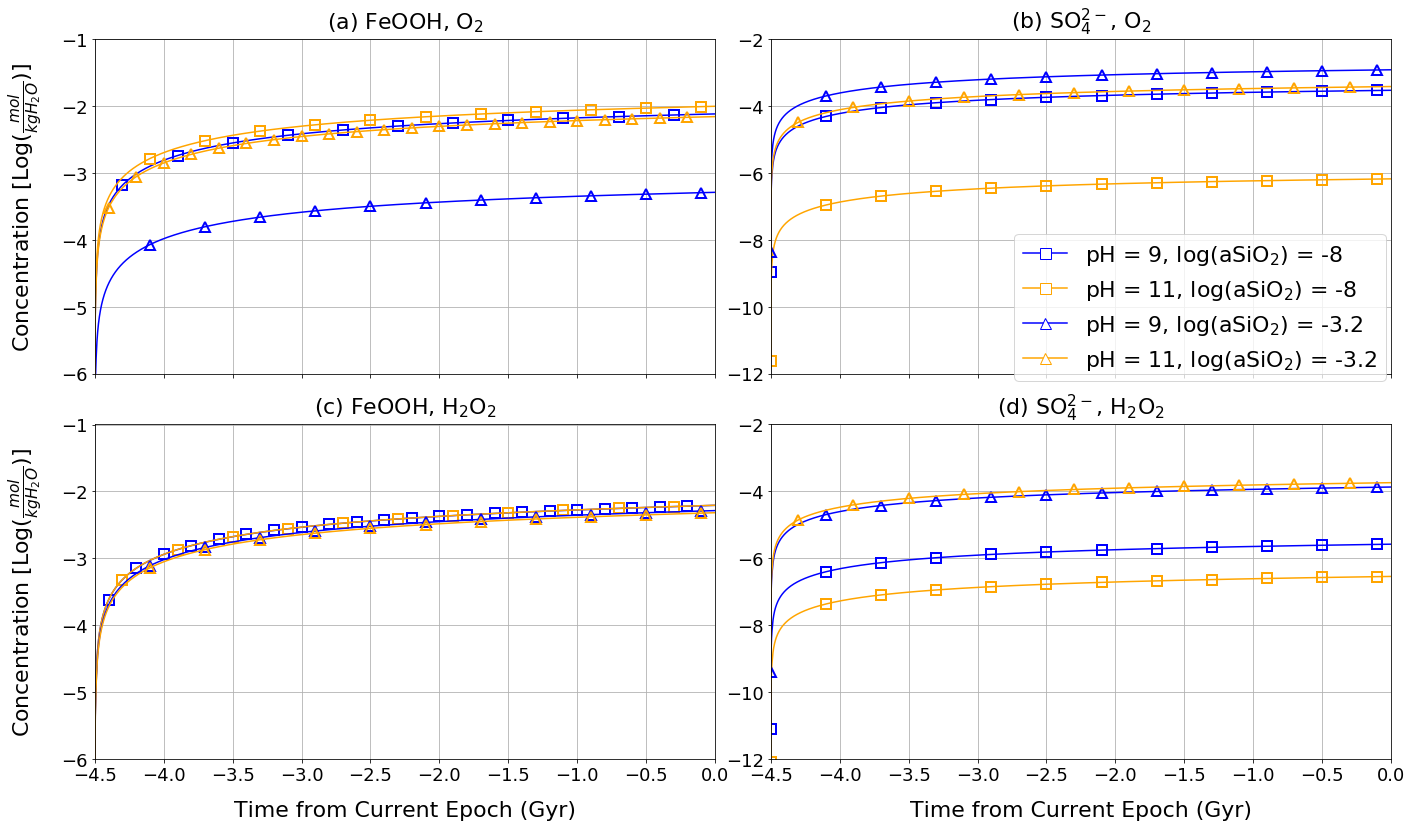}}
\caption{Concentration of FeOOH (a,c) and SO$_4^{2-}$ (b,d) produced by oxidation with O$_2$ (a,b) and H$_2$O$_2$ (c,d) over time for pH 9 and log(aSiO$_2$) = -8 (blue squares), pH 11 and log(aSiO$_2$) = -8 (orange squares), pH 9 and log(aSiO$_2$) = -3.2 (blue triangles), and pH 11 and log(aSiO$_2$) = -3.2 (orange triangles). Note the y-axes scales are different for FeOOH than for SO$_4^{2-}$ in order to better see contrasts between different sets of conditions. For all considered conditions except oxidation by O$_2$ at pH 9 and log(aSiO$_2$) = -3.2, ferrous iron oxidation is strongly favored over sulfide oxidation.}
\label{figone}
\end{figure}
 

\subsection{Case III: Oxidation of Minerals}

In our third case, we assume that reductants are present as minerals in the rocky core, and that all ocean water can circulate through the core such that any O$_2$ and H$_2$O$_2$ present is consumed. To determine how the oxidizing power of O$_2$ and H$_2$O$_2$ are distributed between sulfur-bearing minerals and iron-bearing minerals, we first estimate the relative abundances of these minerals in the core, given the simplified mineralogical model of Enceladus' core from \cite{waite2017cassini} (supplemental material) and reproduced here in Table 3. We consider both a core composed of reduced hydrous rock (RHR) and a core composed of oxidized hydrous rock (OHR), the two models for which the mass of rock in the interior is provided. We take greenalite (Gre) and magnetite (Mag) to be the iron-bearing mineral that will react with oxidants in a RHR or OHR core, respectively, by:

 \begin{table}[h]
    \small
\caption{Mineralogical model of Enceladus' core, modified from Tables S6 and S7 of \cite{waite2017cassini} supplemental material. In case III, we consider iron oxidation in greenalite or magnetite, and sulfur oxidation in pyrrhotite. Chrysotile and talc/saponite are assumed to make up the remainder of the core volume and are shown here for completeness.}
    \begin{tabular}{l c c c}
	    \hline
	   Rock type: & & Reduced & Oxidized \vspace{-2mm}\\
	    & & hydrous rock &  hydrous rock \vspace{-2mm}\\
	    Abbreviation: & & RHR & OHR \\
	    \hline
	    Mineral & Formula & Wt.\% \hspace{2mm} Vol.\% & Wt.\% \hspace{2mm} Vol.\% \\
	    \hline
	    Chrysotile & Mg$_3$Si$_2$O$_5$(OH)$_4$ & 48.86 \hspace{3mm} 57.17 & 27.53 \hspace{3mm} 32.30 \\
	    Greenalite & Fe$_3$Si$_2$O$_5$(OH)$_4$ & 29.75 \hspace{3mm} 27.59 & - \hspace{9mm} - \\
	    Magnetite & Fe$_3$O$_4$ & - \hspace{9mm} - & 19.11 \hspace{3mm} 11.04 \\
	    Pyrrhotite & Fe$_{0.875}$S & 19.27 \hspace{3mm} 12.49 & 19.87 \hspace{3mm} 12.92 \\
	    Talc/Saponite & Mg$_3$Si$_4$O$_{10}$(OH)$_2$ & 2.12 \hspace{5mm} 2.76 & 33.49 \hspace{3mm} 43.74 \\
\hline
    	
	\end{tabular}
\end{table}

\begin{linenomath*}
\begin{equation} 
4\textrm{Fe}_3\textrm{Si}_2\textrm{O}_5(\textrm{OH})_4 + 3\textrm{O}_2 \rightarrow 12\textrm{FeOOH} + 8\textrm{SiO}_2 + 2\textrm{H}_2\textrm{O},
 \end{equation} 
\begin{equation}
2\textrm{Fe}_3\textrm{Si}_2\textrm{O}_5(\textrm{OH})_4 + 3\textrm{H}_2\textrm{O}_2  \rightarrow 6\textrm{FeOOH} + 4\textrm{SiO}_2 + 4\textrm{H}_2\textrm{O},
 \end{equation}
 \begin{equation}
4\textrm{Fe}_3\textrm{O}_4 + \textrm{O}_2 +  6\textrm{H}_2\textrm{O}\rightarrow 12\textrm{FeOOH},
\end{equation}
 \begin{equation}
2\textrm{Fe}_3\textrm{O}_4 + \textrm{H}_2\textrm{O}_2 + 2\textrm{H}_2\textrm{O} \rightarrow 6\textrm{FeOOH}
\end{equation}
\end{linenomath*} 

\noindent and pyrrhotite (Po, Fe$_{0.875}$S) to be the sulfur-bearing mineral that will react with oxidants in both RHR and OHR via:

\begin{linenomath*}
\begin{equation} 
8\textrm{Fe}_{0.875}\textrm{S} + 11.5\textrm{H}_2\textrm{O} + 17.25\textrm{O}_2 \rightarrow 16\textrm{H}^+ + 8\textrm{SO}_4^{2-} + 7\textrm{FeOOH},
 \end{equation} 
\begin{equation}
8\textrm{Fe}_{0.875}\textrm{S} + 34.5\textrm{H}_2\textrm{O}_2  \rightarrow 16\textrm{H}^+ + 8\textrm{SO}_4^{2-} + 23\textrm{H}_2\textrm{O} + 7\textrm{FeOOH}.
 \end{equation}
 \end{linenomath*} 
 
 \begin{figure}[ht!]:
\centerline{\includegraphics[height=3in]{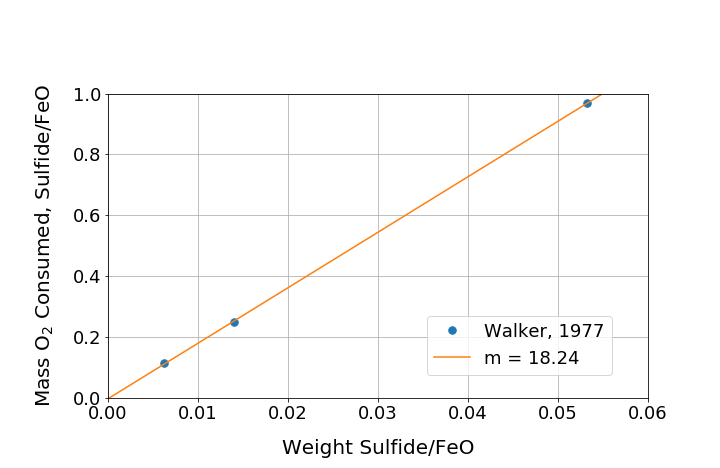}}
\caption{Ratio of O$_2$ consumed per mass sulfide to O$_2$ consumed per mass FeO, versus weight ratio of sulfide to FeO for rocks on Earth from \cite{walker1977atmospherel}. The derived slope of this linear relationship, m = 18, is equal to the ratio of sulfide oxidation to FeO oxidation.}
\label{figone}
\end{figure}

 For both the RHR and OHR interiors, we find the molar ratio of the iron-bearing mineral (greenalite or magnetite) to pyrrhotite to be 1:3. Along with relative abundances, we also consider how quickly iron-bearing minerals oxidize compared with sulfur-bearing minerals. We take the weathering rates for FeO and FeS type minerals by O$_2$ on Earth from \cite{walker1977atmospherel}, and compare the mass of O$_2$ consumed per mass sulfide with the mass of O$_2$ consumed per mass FeO, for a given weight percent ratio of sulfide to FeO, in the rock (Figure 10). Because these weathering rates were derived from a lower pH system than the Enceladus ocean, our extrapolation is missing a pH dependence that could shift the relative rates of iron and sulfur mineral oxidation. However, without weathering data from higher pH systems, we will not consider pH dependence in the current work as it is simply a first attempt to examine how this system might behave, and we emphasize the need for future studies to constrain these rates in alkaline pH environments on Earth.
 
 We find that, for these rocks, sulfide consumes O$_2$ 18 times faster per unit mass. We extrapolate this relationship to the weight percent ratio of sulfide to FeO in RHR and OHR rocks on Enceladus, and find that oxidation of sulfide occurs 2.7 times faster than oxidation of FeO in RHR, and 8.1 times faster in OHR, per unit mole. Assuming the relative oxidation rates for H$_2$O$_2$ are comparable, we compute the fraction of O$_2$ and H$_2$O$_2$ consumed by each rock type via:
 
 \begin{linenomath*}
\begin{equation} 
f_{Gre} = \Big(2.7(\frac{N_{Po,R}}{N_{Gre}}) + 1\Big)^{-1} = 0.11,
\end{equation}
\begin{equation}
f_{Mag} = \Big(8.1(\frac{N_{Po,O}}{N_{Mag}}) + 1\Big)^{-1} = 0.04,
\end{equation}
\begin{equation}
    f_{Po,R} = 1 - f_{Gre} = 0.89,
\end{equation}
\begin{equation}
    f_{Po,O} = 1 - f_{Mag} = 0.96
\end{equation}
\end{linenomath*}  
 
\noindent where N is the number of moles of a given mineral, and the subscripts $R$ or $O$ in $N_{Po}$ and $f_{Po}$ represent either RHR or OHR, respectively. Given the stoichiometry of reactions 29-34, we calculate the amounts of FeOOH and SO$_4^{2-}$ produced through the oxidation of each of these minerals by O$_2$ and H$_2$O$_2$. The results are shown in Figures 11 and 12, with equivalent figures if $^{40}$K decay does not act a source of oxidants (Scenario 3, see Table 1) provided in Appendix B.

Unlike Case II, in which all reductants were in aqueous form (Section 4.2), we find that the relative production of SO$_4^{2-}$ compared to FeOOH is much closer in all cases, for both oxidation by O$_2$ and H$_2$O$_2$, due to the preference for oxidation of FeS rather than FeO. For oxidation by O$_2$, we calculate an upper limit on [FeOOH] of 2.0-2.2 mmol/kg H$_2$O, and an upper limit on [SO$_4^{2-}$] of 1.0-1.1 mmol/kg H$_2$O. For oxidation by H$_2$O$_2$, we calculate an upper limit on [FeOOH] of 1.0-1.1 mmol/kg H$_2$O, and an upper limit on [SO$_4^{2-}$] of 0.52-0.56 mmol/kg H$_2$O.

\begin{figure}[ht!]:
\centerline{\includegraphics[height=3in]{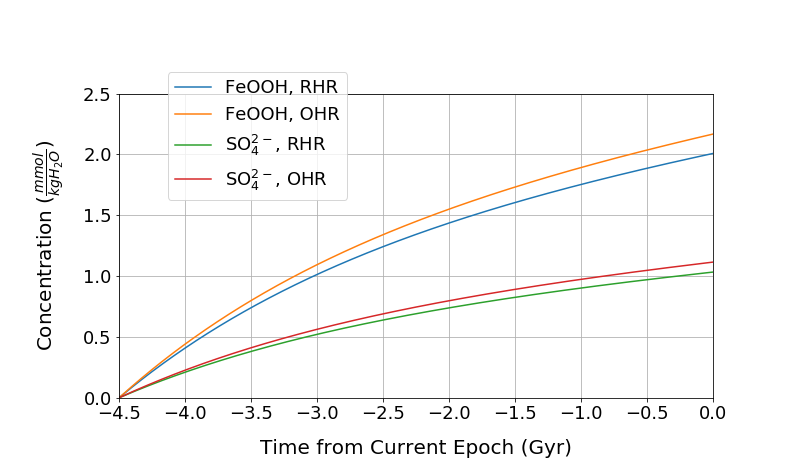}}
\caption{Concentration of FeOOH (blue for RHR, orange for OHR) and SO$_4^{2-}$ (green for RHR, red for OHR) produced by oxidation with O$_2$ over time.}
\label{figone}
\end{figure}
\begin{figure}
\centerline{\includegraphics[height=3in]{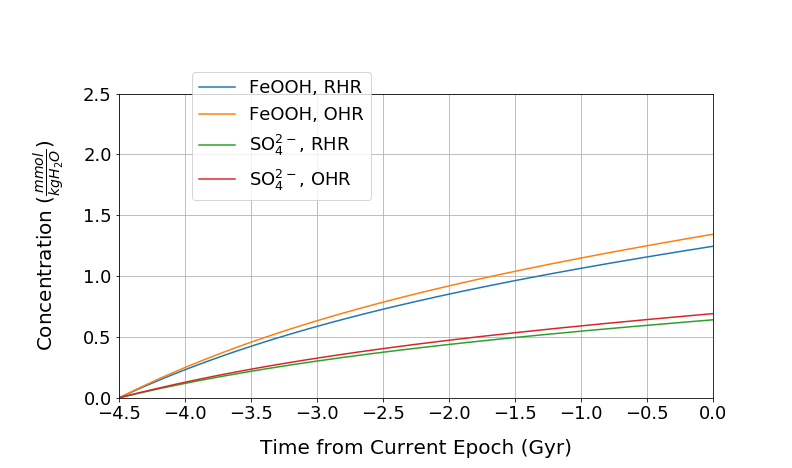}}
\caption{Concentration of FeOOH (blue for RHR, orange for OHR) and SO$_4^{2-}$ (green for RHR, red for OHR) produced by oxidation with H$_2$O$_2$ over time}
\label{figone}
\end{figure}


\section{Compatibility with the Available Data}
To determine which of the three cases outlined in this paper (see Table 1) may be most appropriate, we can examine how the oxidant concentrations predicted in each case compare with available data from Cassini. In Case I, where reactive reductants in the ocean are either produced slowly or are non-existent, the resulting oxidant concentrations are 2.5 mmol O$_2$/kg H$_2$O or 3.1 mmol H$_2$O$_2$/kg H$_2$O when oxidants are formed in the ice shell and as a result of $^{40}$K decay (Scenarios 1 and 2, Table 1), and 1.0 mmol O$_2$/kg H$_2$O or 1.6 mmol H$_2$O$_2$/kg H$_2$O when the ice shell is the only oxidant source (Scenario 3, Table 1). Given that the inferred concentration of CO$_2$ in the ocean from \cite{waite2017cassini} is between $7 \times 10^{-2}$ and $1 \times 10^{-4}$ mmol/kg H$_2$O, this implies that O$_2$ or H$_2$O$_2$  should be $\sim$10 to 30,000 times more abundant than CO$_2$ in the ocean. Because CO$_2$ was detected without ambiguity by INMS (\cite{waite2017cassini}), O$_2$ concentrations this high would also have been detected. Measuring H$_2$O$_2$ would have been more complicated because it is highly reactive, making it difficult to detect in situ. Although \cite{newman2007hydrogen} reported an  H$_2$O$_2$ abundance of $0.05-0.5\%$ with respect to H$_2$O at the tiger stripe region, inferred from Cassini's Visible and Infrared Mapping Spectrometer (VIMS) observations, this is likely all radiolytically produced at the surface, rather than delivered from the ocean. Some fraction of any delivered H$_2$O$_2$ would, however, dissociate into O$_2$ either during plume transport or in the INMS antechamber. Even if this conversion ratio were small (i.e. $\sim1\%$ at pH 9, and as low as $\sim0.001\%$ at pH 11), it would yield O$_2$ abundances comparable to CO$_2$ in the plume. INMS did detect a slight increase in mass 32 signal during the E21 flyby with an upper limit of $\sim0.004\%$, although the signal to noise of this detection was low and could have been due to catalytic conversion of H$_2$O to H$_2$ and O$_2$ on the walls of the antechamber (\cite{waite2017cassini}, supplemental material). 

If we scale the reported $0.004\%$ upper limit on the O$_2$ mixing ratio in the plume gas to CO$_2$, as we did to determine the upper limit on the concentration of H$_2$S in the ocean in Section 4.2.1, this implies an upper limit on the ocean concentration of O$_2$ between $10^{-10}$ and $10^{-6}$ mol/kg H$_2$O. Thus, while O$_2$ or H$_2$O$_2$ may be present at concentrations higher than those predicted in Case II,  observational constraints indicate that their concentrations must still be several orders of magnitude below the concentrations predicted for all scenarios in Case I, if they are present at all. Even if Enceladus is much younger ($\sim 100$ Myr old), the O$_2$ or H$_2$O$_2$ concentrations in Case I would only drop by two orders of magnitude (Figure A.16), and would still be comparable to or greater than the concentration of CO$_2$. At these concentrations O$_2$ should still be detectable. However, at pH = 9, where the H$_2$O$_2$ and CO$_2$ concentrations would be comparable (Figure A.16), the O$_2$ trace left by decomposition of H$_2$O$_2$ may not be detectable depending on the fraction of H$_2$O$_2$ decomposed. A young Enceladus with a pH $\sim 9$ ocean is therefore the only scenario in which Case I cannot be ruled out by constraints from INMS. Even so, aerobic life could still take advantage of the relatively high oxidant fluxes coming from surface ice in Case II and/or Case III if organisms were to reside at the ice-water interface. 

Although Case I may be possible under some conditions, there are several lines of evidence that support a hydrothermally active Enceladus (\cite{hsu2015ongoing}, \cite{waite2017cassini}, \cite{choblet2017powering},  \cite{glein2018geochemistry}), which would both release reductants into the ocean (Case II) and circulate ocean water containing oxidants through the rocky core (Case III). Some support for Case II comes from an ambiguous INMS detection of H$_2$S \citep{magee2017neutral} discussed in Section 4.2.1, which may indicate the presence of sulfide species in the ocean. Both Case II and Case III would also produce SO$_4^{2-}$ which was not detected by CDA (\cite{postberg2009sodium}), but the concentrations of SO$_4^{2-}$ for these cases in Section 4 all fall close to or below the tentative CDA detection limit of 1mM.  The end result of both Case II and Case III are essentially the same (O$_2$ or H$_2$O$_2$ converted to SO$_4^{2-}$ or ferric oxyhydroxides), with the primary difference being the final relative abundances of sulfates versus ferric oxyhydroxides and the energy they could provide (see Section 6). 

With current observational data it is impossible to determine whether one process may dominate over the other (i.e., whether aqueous reductants consume oxidants before they can be circulated through the rocky core, or vice versa), or whether both are occurring to a comparable extent. A rigorous treatment of hydrothermal geochemistry in the rocky core, incorporating the results of \cite{choblet2017powering}, is therefore a necessary next step in determining how these two processes may compete with one another and change the final abundances of oxidants in the ocean. Indeed, if circulation occurs deep enough in the core, radiolysis of water caused by the decay of radionuclides in the core (\cite{lin2005radiolytic}, \cite{lin2005yield}, \cite{bouquet2017alternative}) could act as another source of oxidants to the ocean which could also be incorporated into such a model. Future measurements of sulfur species could also reveal the extent to which oxidation of aqueous reductants versus seafloor/core minerals is occurring. A robust H$_2$S detection, or a detection of ferrous iron, could be used to constrain the concentrations of reduced iron and sulfide species in the ocean and thus the capacity for Case II oxidation to occur. Detection and determination of the ratios of sulfur isotopes in sulfate could also be used not only to confirm the presence of sulfate in the ocean, but also to potentially distinguish between sulfate made from aqueous oxidation of sulfides vs. sulfate that may have formed from sulfur dioxide or elemental sulfur that could have been present in the material from which Enceladus formed. 


\section{Energy Availability for Metabolic Redox Reactions}

To determine whether the abundances of oxidants constrained in Section 4 could be sufficient to provide metabolic energy for life, we calculate chemical affinities for a set of redox reactions, per mole of limiting reactant. These reactions, listed in Tables 4-5 and adapted from \cite{zolotov2004model} and \cite{amend2011catabolic}, encompass many of the major metabolisms used by diverse bacteria and archaea in Earth's oceans and other inhabited environments. We have included both aerobic and anaerobic organic matter oxidation to represent the possibility for life to oxidize the massive organic compounds (HMOC's) detected in the plume (\cite{postberg2018macromolecular}). As an analog for these compounds, we use kerogen compounds with 404, 415, and 515 carbon atoms described in  \cite{helgeson2009chemical}, which well-represent HMOC properties constrained by \cite{postberg2018macromolecular}. Because oxidation of these three kerogens have similar equilibrium constants, the resulting chemical affinities and energy fluxes available from these reactions are similar and we have thus reported the average values produced by all three of these compounds in Tables 4 and 5, with an example reaction for oxidation of kerogen with 415 carbon atoms written out explicitly.

With the exception of sulfide and iron oxidation (reactions 2 and 3 in Tables 4 and 5), we have assumed that all of the metabolic reactions listed in Tables 4 and 5 are kinetically inhibited from proceeding abiotically at low temperatures, for the reasons discussed in Section 4. In biology, however, these could proceed, as biocatalysts (i.e. enzymes) can lower the the activation energy barrier. 

Chemical affinity serves as a measure of disequilibrium for a given reaction; the greater the magnitude of the chemical affinity, the further out of equilibrium the reaction is (\cite{shock2010potential}, \cite{waite2017cassini}). Chemotrophic life takes advantage of this disequilibrium, extracting energy from the environment to drive the system toward equilibrium (\cite{mccollom1997geochemical}, \cite{hoehler2007follow}). The affinity, $A$, can be calculated via:

 \begin{linenomath*}
\begin{equation}
 A = ln(10)RT(\textrm{log}K - \textrm{log}Q)
\end{equation}
\end{linenomath*}

\noindent where $R$ is the gas constant and $T$ is the temperature of the ocean assumed to be 273 K. The log$K$ - log$Q$ term in the equation measures the extent to which the activities of the reaction species are out of equilibrium. We calculate equilibrium constants, $K$, from the CHNOSZ package \citep{dick2019chnosz} at T = 0$^{\circ}$C and P = 73 bar, the approximate pressure at the seafloor \citep{glein2018geochemistry}. We calculate the reaction quotient $Q$ for each reaction from the thermodynamic activities---or effective concentrations---of oxidants O$_2$, H$_2$O$_2$ and SO$_4^{2-}$ and reductants HS$^-$ and Fe$^{2+}$ constrained in Section 4, along with those of H$_2$, CO$_2$ and CH$_4$ derived from INMS detections in \cite{waite2017cassini}. Since FeOOH and Fe$_{0.875}$S are both minerals and kerogen is a solid, their thermodynamic activity is 1 if the minerals are present as pure phases. In Case III we assume the same activities for HS$^-$ and Fe$^{2+}$ predicted in Case II, as it is likely that the two processes considered in these cases are occurring simultaneously if there is hydrothermal activity on Enceladus. If the concentrations of these reductants were lower, that would only push the reactions they are used in further out of equilibrium, such that the affinities and energy fluxes available would be even higher than the numbers we are reporting. For Case I, we do not consider reactions involving iron or sulfur species, as we have assumed that these reductants are not present.

The chemical affinities for our selected reactions, calculated for each case and scenario outlined in Table 1 and for a young (100 Myr) and old (4.5 Gyr) Enceladus, are shown in Table 4. We use methanogenesis, for which the affinity is less uncertain because it is constrained entirely by observations, as a basis for comparison with all of the other investigated reactions. To determine whether these reactions could be used as metabolisms, we compare their affinities to $\Delta G_{min}$, the minimum amount of energy required to convert ADP to ATP and therefore be used by life on Earth, which has been measured to be as low as 20 kJ/mol for laboratory cultures and 10 kJ/mol for cells in nature (\citep{hoehler2001apparent}, \citep{hoehler2004biological}). Methanogenesis clearly meets these requirements for both pH 9 and 11. The affinities for the chosen aerobic reactions are also well above this requirement, even for the Case II values for which the concentrations of O$_2$ and H$_2$O$_2$ are very low (Section 4.2). Although we have assumed aerobic metabolism is not possible in Case III (Section 4.3), life could likely still use some of the O$_2$ and/or H$_2$O$_2$ flux produced in the ocean even if these oxidants were also reacting with minerals. Aerobic reactions could therefore likely support life on Enceladus even if hydrothermal activity were consuming a large portion of the oxidant flux at the same time.

 \begin{sidewaystable}[h]
    \setlength{\tabcolsep}{0.5pt}
    \caption{Range of chemical affinities for metabolic reactions. The "Primary Model" is a 4.5 Gyr old Enceladus with both an ice shell and ocean oxidant source, for which corresponding plots are provided throughout the main text. Values are also shown for a young (100 Myr old) Enceladus, for which corresponding figures are provided in Appendix A, and for Scenario 3 in which only the ice shell provides oxidants, for which corresponding figures are provided in Appendix B. For each of these three possible models, we have shown values for all three of the cases outlined in Table 1. Values for all reactions have been rounded to the nearest ten. Values reported with a range correspond to metabolisms in which different ocean conditions changed the activity of one or more species and resulted in an affinity that varied by more than the rounding error.}
	\resizebox{1.2\textwidth}{!}{\begin{tabular}{|l|ccc|ccc|ccc|}
	    \hline
	    \multicolumn{1}{|c}{\textbf{Metabolic Process}} & \multicolumn{9}{|c|}{\textbf{Affinity (kJ mol$^{-1}$)}} \\
	    \hline
	    \multicolumn{1}{|l|}{Methanogenesis:} & \multicolumn{9}{c|}{ }\\
	    \multicolumn{1}{|l|}{ 4H$_{2}$ + CO$_{2} \rightarrow$ CH$_{4}$ + 2H$_{2}$O} & \multicolumn{9}{c|}{90 $\pm$ 30} \\
	    \hline
	    
	     \multicolumn{1}{|l|}{\cellcolor{gray}} & \multicolumn{3}{c|}{Primary Model} & \multicolumn{3}{c|}{Young Enceladus} & \multicolumn{3}{c|}{No $^{40}$K Oxidant Source} \\
	    \hline
	    Aerobic Reactions & Case I & Case II & Case III  & Case I & Case II & Case III  & Case I & Case II & Case III \\
	    \hline
	    \multicolumn{1}{|l|}{1. Methane Oxidation:} & \multicolumn{3}{c|}{} & \multicolumn{3}{c|}{} & \multicolumn{3}{c|}{} \\
	    \hspace{3mm} $\frac{1}{2}$CH$_4$ + O$_2 \rightarrow$ $\frac{1}{2}$CO$_2$ + H$_2$O & 420 & 360 $\pm$ 10 & NA & 410 & 360 $\pm$ 10 & NA & 410 & 360 $\pm$ 10 & NA \\
	    
	    \hspace{3mm} $\frac{1}{4}$CH$_4$ + H$_2$O$_2 \rightarrow \frac{1}{4}$CO$_2$ + $\frac{6}{4}$H$_2$O & 300 & 250 $\pm$ 40 & NA & 290 & 250 $\pm$ 40 & NA & 300 & 250 $\pm$ 40 & NA \\
    	
	    \multicolumn{1}{|l|}{2. Oxidation of Sulfides} & \multicolumn{3}{c|}{} & \multicolumn{3}{c|}{} & \multicolumn{3}{c|}{} \\
	    \hspace{3mm} $\frac{1}{2}$HS$^-$ + O$_{2} \rightarrow$ $\frac{1}{2}$SO$_{4}^{2-}$ + $\frac{1}{2}$H$^{+}$ & NA & 350 $\pm$ 10 & NA & NA & 350 $\pm$ 10 & NA & NA & 350 $\pm$ 10 & NA \\  
	    
	    \hspace{3mm} $\frac{1}{4}$HS$^-$ + H$_2$O$_{2} \rightarrow$ $\frac{1}{4}$SO$_{4}^{2-}$ + $\frac{1}{4}$H$^{+}$ +H$_2$O & NA &  220 $\pm$ 10  & NA & NA & 220 $\pm$ 10 & NA & NA & 220 $\pm$ 10 & NA \\ 
	    
	    \multicolumn{1}{|l|}{3. Oxidation of Iron Sulfide} & \multicolumn{3}{c|}{} & \multicolumn{3}{c|}{} & \multicolumn{3}{c|}{} \\   
	    \hspace{3mm} $\frac{16}{31}$Fe$_{0.875}$S + O$_{2}$ + $\frac{2}{31}$H$_{2}$O $\rightarrow$ $\frac{14}{31}$Fe$^{2+}$ + $\frac{16}{31}$SO$_{4}^{2-}$ +  $\frac{4}{31}$H$^{+}$ & NA & 350 $\pm$ 10 & NA & NA & 350 $\pm$ 10 & NA & NA & 350 $\pm$ 10 & NA \\
	   
	    \hspace{3mm} $\frac{8}{31}$Fe$_{0.875}$S + H$_2$O$_{2}$ $\rightarrow$ $\frac{7}{31}$Fe$^{2+}$ + $\frac{8}{31}$SO$_{4}^{2-}$ +  $\frac{2}{31}$H$^{+}$ + $\frac{30}{31}$H$_2$O & NA & 220 $\pm$ 10 & NA & NA & 220 $\pm$ 10 & NA & NA & 220 $\pm$ 10 & NA\\

	    \multicolumn{1}{|l|}{4. Hydrogen Oxidation:} & \multicolumn{3}{c|}{} & \multicolumn{3}{c|}{} & \multicolumn{3}{c|}{} \\ 	   
	    \hspace{3mm} 2H$_{2}$ + O$_{2} \rightarrow$ 2H$_{2}$O & \hspace{4mm} 460 $\pm$ 10 \hspace{4mm} & 400 $\pm$ 10 & NA & \hspace{4mm} 450 $\pm$ 10 \hspace{4mm} & \hspace{1mm} 400 $\pm$ 10 \hspace{1mm} & NA & \hspace{4mm} 460 $\pm$ 10 \hspace{4mm} & 400 $\pm$ 10 & NA \\
	    
	    \hspace{3mm} H$_{2}$ + H$_2$O$_{2} \rightarrow$ 2H$_{2}$O &  \hspace{0.5mm} 320 $\pm$ 10 \hspace{0.5mm} & 250 $\pm$ 10  & NA & 310 $\pm$ 10 & 250 $\pm$ 10 & NA & 320 $\pm$ 10 & 240 $\pm$ 10 & NA \\

	    \multicolumn{1}{|l|}{5. Aerobic Organic Matter Oxidation:} & \multicolumn{3}{c|}{} & \multicolumn{3}{c|}{} & \multicolumn{3}{c|}{} \\ 	   	   
	    \hspace{3mm} $\frac{2}{1157}$C$_{415}$H$_{698}$O$_{22}$ + O$_2$ $\rightarrow$ $\frac{830}{1157}$CO$_2$ + $\frac{698}{1157}$H$_2$O & 440 $\pm$ 10 &  370 $\pm$ 10 & NA & 430 $\pm$ 10 & 370 $\pm$ 10 & NA & 440 $\pm$ 10 & 370 $\pm$ 10 & NA \\
	    
	    $\frac{1}{1157}$C$_{415}$H$_{698}$O$_{22}$ + H$_2$O$_2$ $\rightarrow$ $\frac{415}{1157}$CO$_2$ + $\frac{1506}{1157}$H$_2$O & 310 $\pm$ 10 & 240 $\pm$ 20 & NA & 300 $\pm$ 10 & 240 $\pm$ 20 & NA & 310 $\pm$ 10 & 240 $\pm$ 20 & NA \\
	    
    	\hline
    	Anaerobic Reactions & Case I & Case II & Case III  & Case I & Case II & Case III  & Case I & Case II & Case III \vspace{-1mm}\\
    	\hline

	    \multicolumn{1}{|l|}{1. Sulfate Reduction:} & \multicolumn{3}{c|}{} & \multicolumn{3}{c|}{} & \multicolumn{3}{c|}{} \\ 	   	   
	     \hspace{3mm} 4H$_{2}$ + SO$_{4}^{2-}$ + H$^{+} \rightarrow$ 4H$_{2}$O +  HS$^{-}$ & NA & 100 $\pm$ 40 & \hspace{3mm} 110 $\pm$ 30 \hspace{3mm} & NA & 90 $\pm$ 40 & \hspace{3mm} 90 $\pm$ 30 \hspace{3mm} & NA & 100 $\pm$ 40 & \hspace{3mm} 100 $\pm$ 30 \hspace{3mm} \\ 

	    \multicolumn{1}{|l|}{2. Anaerobic Oxidation of Methane:} & \multicolumn{3}{c|}{} & \multicolumn{3}{c|}{} & \multicolumn{3}{c|}{} \\	 	    
	    \hspace{3mm} CH$_{4}$ + SO$_{4}^{2-}$ + H$^{+} \rightarrow$ CO$_{2}$ + HS$^-$ + 2H$_{2}$O & NA & 10 $\pm$ 10 & 20 & NA & 0 $\pm$ 10 & 10 & NA & 10 $\pm$ 10 & 20 \\ 

	    \multicolumn{1}{|l|}{3. Reduction of Ferric Iron:} & \multicolumn{3}{c|}{} & \multicolumn{3}{c|}{} & \multicolumn{3}{c|}{} \\	   
	    \hspace{3mm} $\frac{1}{2}$H$_2$ + FeOOH + 2H$^+ \rightarrow$ 2H$_2$O + Fe$^{2+}$ & NA & 20 $\pm$ 10 & 10 $\pm$ 10 & NA & 20 $\pm$ 10 & 10 $\pm$ 10 & NA & 20 $\pm$ 10 & 10 $\pm$ 10 \\ 

	    \multicolumn{1}{|l|}{4. Anaerobic Organic Matter Oxidation:} & \multicolumn{3}{c|}{} & \multicolumn{3}{c|}{} & \multicolumn{3}{c|}{} \\		    
        \hspace{3mm} $\frac{4}{1157}$C$_{415}$H$_{698}$O$_{22}$ + SO$_4^{2-}$ + H$^+$ $\rightarrow$ $\frac{1660}{1157}$CO$_2$ + HS$^-$ + $\frac{1396}{1157}$H$_2$O & NA & 70 $\pm$ 20 & 80 $\pm$ 10 & NA & 60 $\pm$ 20 & 70 $\pm$ 10 & NA & 70 $\pm$ 20 & 70 $\pm$ 10 \\ 
        
        \hspace{3mm} $\frac{1}{2314}$C$_{415}$H$_{698}$O$_{22}$ + FeOOH + 2H$^+$ $\rightarrow$ $\frac{415}{2314}$CO$_2$ + Fe$^{2+}$ + $\frac{1910}{1157}$H$_2$O & NA & 10 $\pm$ 10 & 10 $\pm$ 10 & NA & 10 $\pm$ 10 & 10 $\pm$ 10 & NA & 10 $\pm$ 10 & 10 $\pm$ 10 \\ 
        \hline 
	\end{tabular}}
\end{sidewaystable}

 \begin{sidewaystable}[h]
    \setlength{\tabcolsep}{0.5pt}
    \caption{Range of energy fluxes for metabolic reactions. The "Primary Model" is a 4.5 Gyr old Enceladus with both an ice shell and ocean oxidant source, for which corresponding plots are provided throughout the main text. Values are also shown for a young (100 Myr old) Enceladus, for which corresponding plots are provided in Appendix A, and for Scenario 3 in which only the ice shell provides oxidants, for which corresponding plots are provided in Appendix B. For each of these possible models, we have shown values for all three of the cases outlined in Table 1. Values for aerobic reactions have been rounded to the nearest ten, and values for anaerobic reactions have been rounded to the nearest one. Values reported with a range correspond to metabolisms in which different ocean conditions changed the activity of one or more species and resulted in an energy flux that varied by more than the rounding error.}
	\resizebox{1.1\textwidth}{!}{\begin{tabular}{|l|ccc|ccc|ccc|}
	    \hline
	    \multicolumn{1}{|c}{\textbf{Metabolic Process}} & \multicolumn{9}{|c|}{\textbf{Energy Flux (kJ s$^{-1}$)}} \\
	    \hline
	    \multicolumn{1}{|l|}{Methanogenesis:} & \multicolumn{9}{c|}{ }\\
	    \multicolumn{1}{|l|}{ 4H$_{2}$ + CO$_{2} \rightarrow$ CH$_{4}$ + 2H$_{2}$O} & \multicolumn{9}{c|}{10,300 $\pm$ 7,700} \\
	    \hline
	    
	     \multicolumn{1}{|l|}{\cellcolor{gray}} & \multicolumn{3}{c|}{Primary Model} & \multicolumn{3}{c|}{Young Enceladus} & \multicolumn{3}{c|}{No $^{40}$K Oxidant Source} \\
	    \hline
	    Aerobic Reactions & Case I & Case II & Case III  & Case I & Case II & Case III  & Case I & Case II & Case III \\
	    \hline
	    \multicolumn{1}{|l|}{1. Methane Oxidation:} & \multicolumn{3}{c|}{} & \multicolumn{3}{c|}{} & \multicolumn{3}{c|}{} \\
	    \hspace{3mm} $\frac{1}{2}$CH$_4$ + O$_2 \rightarrow$ $\frac{1}{2}$CO$_2$ + H$_2$O & 110 & 90 & NA & 100 & 90 & NA & 80 & 70 & NA \\
	    
	    \hspace{3mm} $\frac{1}{4}$CH$_4$ + H$_2$O$_2 \rightarrow \frac{1}{4}$CO$_2$ + $\frac{6}{4}$H$_2$O & 110 & 90 $\pm$ 10 & NA & 110 & 90 $\pm$ 10 & NA & 90 & 70 $\pm$ 10 & NA \\
    	
	    \multicolumn{1}{|l|}{2. Oxidation of Sulfides} & \multicolumn{3}{c|}{} & \multicolumn{3}{c|}{} & \multicolumn{3}{c|}{} \\
	    \hspace{3mm} $\frac{1}{2}$HS$^-$ + O$_{2} \rightarrow$ $\frac{1}{2}$SO$_{4}^{2-}$ + $\frac{1}{2}$H$^{+}$ & NA & 90 & NA & NA & 90 & NA & NA & 60 & NA \\  
	    
	    \hspace{3mm} $\frac{1}{4}$HS$^-$ + H$_2$O$_{2} \rightarrow$ $\frac{1}{4}$SO$_{4}^{2-}$ + $\frac{1}{4}$H$^{+}$ +H$_2$O & NA & 80 & NA & NA & 80 & NA & NA & 70 & NA \\ 
	    
	    \multicolumn{1}{|l|}{3. Oxidation of Iron Sulfide} & \multicolumn{3}{c|}{} & \multicolumn{3}{c|}{} & \multicolumn{3}{c|}{} \\   
	    \hspace{3mm} $\frac{16}{31}$Fe$_{0.875}$S + O$_{2}$ + $\frac{2}{31}$H$_{2}$O $\rightarrow$ $\frac{14}{31}$Fe$^{2+}$ + $\frac{16}{31}$SO$_{4}^{2-}$ +  $\frac{4}{31}$H$^{+}$ & NA & 90 & NA & NA & 90 & NA & NA & 60 & NA \\
	   
	    \hspace{3mm} $\frac{8}{31}$Fe$_{0.875}$S + H$_2$O$_{2}$ $\rightarrow$ $\frac{7}{31}$Fe$^{2+}$ + $\frac{8}{31}$SO$_{4}^{2-}$ +  $\frac{2}{31}$H$^{+}$ + $\frac{30}{31}$H$_2$O & NA & 80 $\pm$ 10 & NA & NA & 80 & NA & NA & 70 & NA\\

	    \multicolumn{1}{|l|}{4. Hydrogen Oxidation:} & \multicolumn{3}{c|}{} & \multicolumn{3}{c|}{} & \multicolumn{3}{c|}{} \\ 	   
	    \hspace{3mm} 2H$_{2}$ + O$_{2} \rightarrow$ 2H$_{2}$O & \hspace{4mm} 120 \hspace{4mm} & 100 & NA & \hspace{4mm} 110 \hspace{4mm} & \hspace{1mm} 100 \hspace{1mm} & NA & \hspace{4mm} 80 \hspace{4mm} & 70 & NA \\
	    
	    \hspace{3mm} H$_{2}$ + H$_2$O$_{2} \rightarrow$ 2H$_{2}$O &  \hspace{0.5mm} 120 \hspace{0.5mm} & 90 & NA & 110 & 90 & NA & 100 & 70 & NA \\

	    \multicolumn{1}{|l|}{5. Aerobic Organic Matter Oxidation:} & \multicolumn{3}{c|}{} & \multicolumn{3}{c|}{} & \multicolumn{3}{c|}{} \\ 	   	   
	    \hspace{3mm} $\frac{2}{1157}$C$_{415}$H$_{698}$O$_{22}$ + O$_2$ $\rightarrow$ $\frac{830}{1157}$CO$_2$ + $\frac{698}{1157}$H$_2$O & 110 & 90 & NA & 110 & 90 & NA & 80 & 70 & NA \\
	    
	    $\frac{1}{1157}$C$_{415}$H$_{698}$O$_{22}$ + H$_2$O$_2$ $\rightarrow$ $\frac{415}{1157}$CO$_2$ + $\frac{1506}{1157}$H$_2$O & 120 & 90 $\pm$ 10 & NA & 110 & 90 $\pm$ 10 & NA & 90 & 70 & NA \\
	    
    	\hline
    	Anaerobic Reactions & Case I & Case II & Case III  & Case I & Case II & Case III  & Case I & Case II & Case III \vspace{-1mm}\\
    	\hline

	    \multicolumn{1}{|l|}{1. Sulfate Reduction:} & \multicolumn{3}{c|}{} & \multicolumn{3}{c|}{} & \multicolumn{3}{c|}{} \\ 	   	   
	     \hspace{3mm} 4H$_{2}$ + SO$_{4}^{2-}$ + H$^{+} \rightarrow$ 4H$_{2}$O +  HS$^{-}$ & NA & 2 $\pm$ 15 & \hspace{3mm} 9 $\pm$ 6 \hspace{3mm} & NA & 2 $\pm$ 14 & \hspace{3mm} 9 $\pm$ 6 \hspace{3mm} & NA & 1 $\pm$ 10 & \hspace{3mm} 6 $\pm$ 5 \hspace{3mm} \\ 

	    \multicolumn{1}{|l|}{2. Anaerobic Oxidation of Methane:} & \multicolumn{3}{c|}{} & \multicolumn{3}{c|}{} & \multicolumn{3}{c|}{} \\	 	    
	    \hspace{3mm} CH$_{4}$ + SO$_{4}^{2-}$ + H$^{+} \rightarrow$ CO$_{2}$ + HS$^-$ + 2H$_{2}$O & NA & 0 $\pm$ 2 & 2$\pm$ 1 & NA & 0 $\pm$ 1 & 1 & NA & 0 $\pm$ 1 & 1 \\ 

	    \multicolumn{1}{|l|}{3. Reduction of Ferric Iron:} & \multicolumn{3}{c|}{} & \multicolumn{3}{c|}{} & \multicolumn{3}{c|}{} \\	   
	    \hspace{3mm} $\frac{1}{2}$H$_2$ + FeOOH + 2H$^+ \rightarrow$ 2H$_2$O + Fe$^{2+}$ & NA & 11 $\pm$ 5 & 3 $\pm$ 3 & NA & 11 $\pm$ 5 & 3 $\pm$ 3 & NA & 8 $\pm$ 5 & 2 $\pm$ 2 \\ 

	    \multicolumn{1}{|l|}{4. Anaerobic Organic Matter Oxidation:} & \multicolumn{3}{c|}{} & \multicolumn{3}{c|}{} & \multicolumn{3}{c|}{} \\		    
        \hspace{3mm} $\frac{4}{1157}$C$_{415}$H$_{698}$O$_{22}$ + SO$_4^{2-}$ + H$^+$ $\rightarrow$ $\frac{1660}{1157}$CO$_2$ + HS$^-$ + $\frac{1396}{1157}$H$_2$O & NA & 1 $\pm$ 7 & 7 $\pm$ 3 & NA & 1 $\pm$ 6 & 6 $\pm$ 2 & NA & 1 $\pm$ 5 & 5 $\pm$ 2 \\ 
        
        \hspace{3mm} $\frac{1}{2314}$C$_{415}$H$_{698}$O$_{22}$ + FeOOH + 2H$^+$ $\rightarrow$ $\frac{415}{2314}$CO$_2$ + Fe$^{2+}$ + $\frac{1910}{1157}$H$_2$O & NA & 6 $\pm$ 7 & 2 $\pm$ 2 & NA & 6 $\pm$ 7 & 2 $\pm$ 2 & NA & 5 $\pm$ 5 & 2 $\pm$ 1 \\ 
        \hline 
	\end{tabular}}
\end{sidewaystable}

Because SO$_4^{2-}$ and ferric oxyhydroxides are much weaker oxidants than O$_2$, even when they are present in large abundances (as in Case II and Case III) the capacity for anaerobic reactions to meet $\Delta G_{min}$ is largely affected by ocean chemistry, the extent to which water-rock reactions could be occurring on Enceladus, and the time period over which oxidant production could be occurring. Case III represents the most favorable set of conditions for anaerobic reactions to meet $\Delta G_{min}$. Thus, if ocean water can percolate through the seafloor, particularly at warmer temperatures where aqueous O$_2$ and H$_2$O$_2$ can readily oxidize reduced iron and sulfur-bearing minerals there, then anaerobic reactions could more feasibly serve as an energy source for life in the ocean. In Case II, where seafloor water-rock reactions serve primarily to produce aqueous reductants in the ocean faster than the production of oxidants, but where minerals are not oxidized directly, the capacity for anaerobic reactions to meet the $\Delta G_{min}$ depends strongly on ocean chemistry, and particularly on pH. In a scenario like Case II in which all O$_2$ or H$_2$O$_2$ abiotically oxidizes iron and sulfur in the ocean before it can be used by life, the pH of the ocean would limit which anaerobic metabolisms could support life. 

Besides the $\Delta G_{min}$, another critical factor that determines whether a given redox coupling could actually support life is the flux of that energy source through time. The magnitude of this energy flux determines how much biomass could be created and supported within Enceladus. We convert our affinity values to energy fluxes by multiplying by the production rate of the limiting reactant, which is the oxidant for all reactions besides methanogenesis. The results are reported in Table 5. For methanogenesis, we assume the flux of H$_2$ in the plume, $1-5 \times 10^9$ mol yr$^{-1}$ \citep{waite2017cassini}, is equal to the production rate of H$_2$ in the ocean. The lower and upper limits on this production rate are reflected in the range of values reported in Table 5. This assumption is valid if the concentration of H$_2$ in the ocean is in steady-state and fractionation in the plume is accounted for. The range of H$_2$ production rates estimated this way is consistent with some values reported from models of serpentinization of Enceladus' rocky core, including \cite{vance2016geophysical} and the upper limit in \cite{steel2017abiotic}. Other reported values range between two orders of magnitude lower (the lower limit reported in \cite{steel2017abiotic}) and four to five orders of magnitude lower \citep{taubner2018biological}, which would lower the energy flux from methanogenesis proportionally. 

We can explore the size of the biosphere that may be supported on Enceladus by comparing our energy flux values to maintenance energy requirements for microbial cells on Earth, or the energy flux required to sustain some amount of biomass (\cite{hoehler2004biological} and \cite{hoehler2013microbial}). Although exact values for maintenance energies are not well constrained, particularly for organisms in nature, we can obtain a general understanding of where the values in Table 5 fall by comparing them to values determined from chemostat cultures ($3.2 \times 10^{-13}$ kJ cell$^{-1}$ day$^{-1}$ for anaerobes and $5.6 \times 10^{-13}$ kJ cell$^{-1}$ day$^{-1}$ for aerobes, \cite{tijhuis1993thermodynamically} and \cite{hoehler2013microbial}). These values could, however, be several orders of magnitude lower for organisms living in natural, and particularly in energy-starved, environments (\cite{morita1997bacteria}, \cite{morita1999h}). We use these values, and the ranges of energy fluxes provided by each reaction in Table 5, to estimate the number of cells that could be supported by these metabolisms in each case (Table 6). For closer comparison to Earth systems, we also calculate theoretical cell densities over a range of possible volumes, with three nominal reference points in which life on Enceladus could be contained: the area of the seafloor directly under the south polar terrain (SPT) where we predict the O$_2$ and H$_2$O$_2$ flux will be highest, the entire seafloor, and the entire ocean volume (Figure 13). For the former two we have considered a range of depths from 1 cm to 20 m, which could encompass microbes in a thin mat at the seafloor or spread down to the \cite{roy2012aerobic} Earth reference point shown on Figure 13. It is possible that microbes could inhabit depths greater than these depending on the temperature in Enceladus' rocky core, but the resulting cell densities would be even smaller than those reference points reported for Earth from \cite{whitman1998prokaryotes}, \cite{christner2006limnological} and \cite{roy2012aerobic}.

It is evident from both Table 6 and Figure 13 that, while methanogenesis clearly dominates as an energy source at least for the high hydrogen production rates that we have used, all of the potential metabolic reactions considered here could still contribute to supporting a biosphere on Enceladus. The cell densities we have calculated for reactions other than methanogenesis are low compared to values measured on Earth (see Figure 13), but could become comparable to these values if smaller volumes were considered. Although ocean chemistry in Case II can have a significant impact on the capacity for anaerobic metabolisms to support life, some anaerobic metabolisms and all aerobic metabolisms are still able to sustain biomass in each set of conditions we have considered. This is significant not only because it supports the potential for diverse metabolisms on Enceladus, but also because of its implications for how life could have emerged there if the availability of redox species was variable throughout Enceladus' history. For instance, if there was less tidal heating of Enceladus' rocky core earlier in its history and H$_2$ availability was more restricted, other redox reactions besides methanogenesis could still provide energy for the development of life. 

 \begin{table}[ht!]
\caption{Total cells that could be supported from our list of metabolic redox reactions, based on maintenance energy requirements for chemostat cell cultures. Cells can be supported by all of our considered aerobic and anaerobic reactions except in some Case II conditions, where the energy fluxes were insufficient to support any life. Thus, the corresponding cell counts are equal to zero.}
\centering
    \hspace*{-1cm}
\resizebox{1.3\textwidth}{!}{\begin{tabular}{|l| c| c| c| }
	\hline
	  & \multicolumn{3}{c|}{\textbf{Total Cells}}  \\
	\hline
    \textbf{Methanogenesis:} & \multicolumn{3}{c|}{7.0e20 - 4.8e21} \\
 	\hline
 	 \cellcolor{gray} & Primary Model & Young Enceladus & No $^{40}$K Oxidant Source \\
    \hline 
 	\textbf{Case I: Buildup of O$_2$ and H$_2$O$_2$} & & &  \\
 	Aerobes: & 1.3e19 - 1.9e19 & 1.2e19 - 1.8e19 & 1.0e19 - 1.5e19 \\
 	\hline
 	\textbf{Case II: Oxidation of Aqueous Reductants} & & & \\
 	Aerobes: & 8.2e18 - 1.6e19 & 8.2e18 - 1.6e19 & 6.6e18 - 1.3e19 \\
 	Anaerobes: & 0 - 4.4e18 & 0 - 4.3e18 & 0 - 3.5e18 \\
 	\hline
 	\textbf{Case III: Oxidation of Minerals} & & &  \\
 	Anaerobes: & 5.5e16 - 4.2e18 & 5.5e16 - 3.8e18 & 4.4e16 - 3.0e18 \\
 	\hline
	\end{tabular}}
\end{table}

If the age of Enceladus is closer to 100 Myr, Tables 4-6 show that the capacity for the ocean to sustain life does not appreciably change from an energetic standpoint. Although the final concentrations of oxidants in the ocean decrease by one to two orders of magnitude (with the exception of the steady-state concentrations of O$_2$ and H$_2$O$_2$ in Case II, which are the same - see Appendix A), the resulting affinities drop by only 10-20 kJ/mol, if at all. This only affects the potential for a few anaerobic reactions to meet the $\Delta G_{min}$, while the affinities for all aerobic and most anaerobic metabolisms still exceed this threshold. Because the production rates of oxidants at present day are the same regardless of the age of Enceladus, the energy fluxes are affected only by affinity and thus are also either unchanged or decreased by only 10 kJ s$^{-1}$ or less. The cell counts resulting from these decreased fluxes would only drop by 10$\%$ or less (Table 6). The production rates, and the resulting energy fluxes and supported cell densities, could be further decreased if the surface at the tiger stripe region was less active, or not active at all earlier in Enceladus' history. However, even if the surface contribution to the O$_2$ and H$_2$O$_2$ production rate was effectively cut off, oxidant production by $^{40}K$ decay would persist as long as there was an ocean present, unless aqueous reductants caused significant interference with oxidant production (Scenario 3). Furthermore, because the rate of radiolytic oxidant production by $^{40}$K decay decreases through time (Appendix C, Figure C.27), the energy flux available from our selected reactions could have been higher earlier in Enceladus' history, even if the contribution from the surface were smaller. In the case of a 4.5 Gyr old Enceladus, the O$_2$ or H$_2$O$_2$ production rates are dominated by $^{40}K$ decay for the first $\sim2.75$ or $1.8$ Gyr, respectively, which would result in higher energy fluxes even if there were no oxidants delivered from the surface during this time. Thus, even for a younger Enceladus and variable surface overturn rates, our chosen redox couplings might still be exploited as energy sources by microbial life in the ocean.

The Enceladus ocean's ability to sustain life similarly does not change appreciably if there is no oxidant production from $^{40}$K decay, a possibility discussed in Section 3. While the total oxidant availability in all three cases is lowered by up to an order of magnitude (Appendix B), the chemical affinities do not drop by more than 10 kJ/mole at most (see Table 4). The energy fluxes are slightly more sensitive to this potential decrease in oxidant availability, as they also depend on the oxidant production rate and would be up to 30 kJ s$^{-1}$ lower for aerobic reactions, and up to $\sim$5 kJ s$^{-1}$ lower for anaerobic reactions (see Table 5). The resulting total number of cells that could be made would thus decrease by 20-30 $\%$. However, as with a younger Enceladus age, this hardly affects the size of the biosphere (i.e. the cell densities) that could be sustained.

\begin{figure}[ht!]
\centerline{\includegraphics[height=3in]{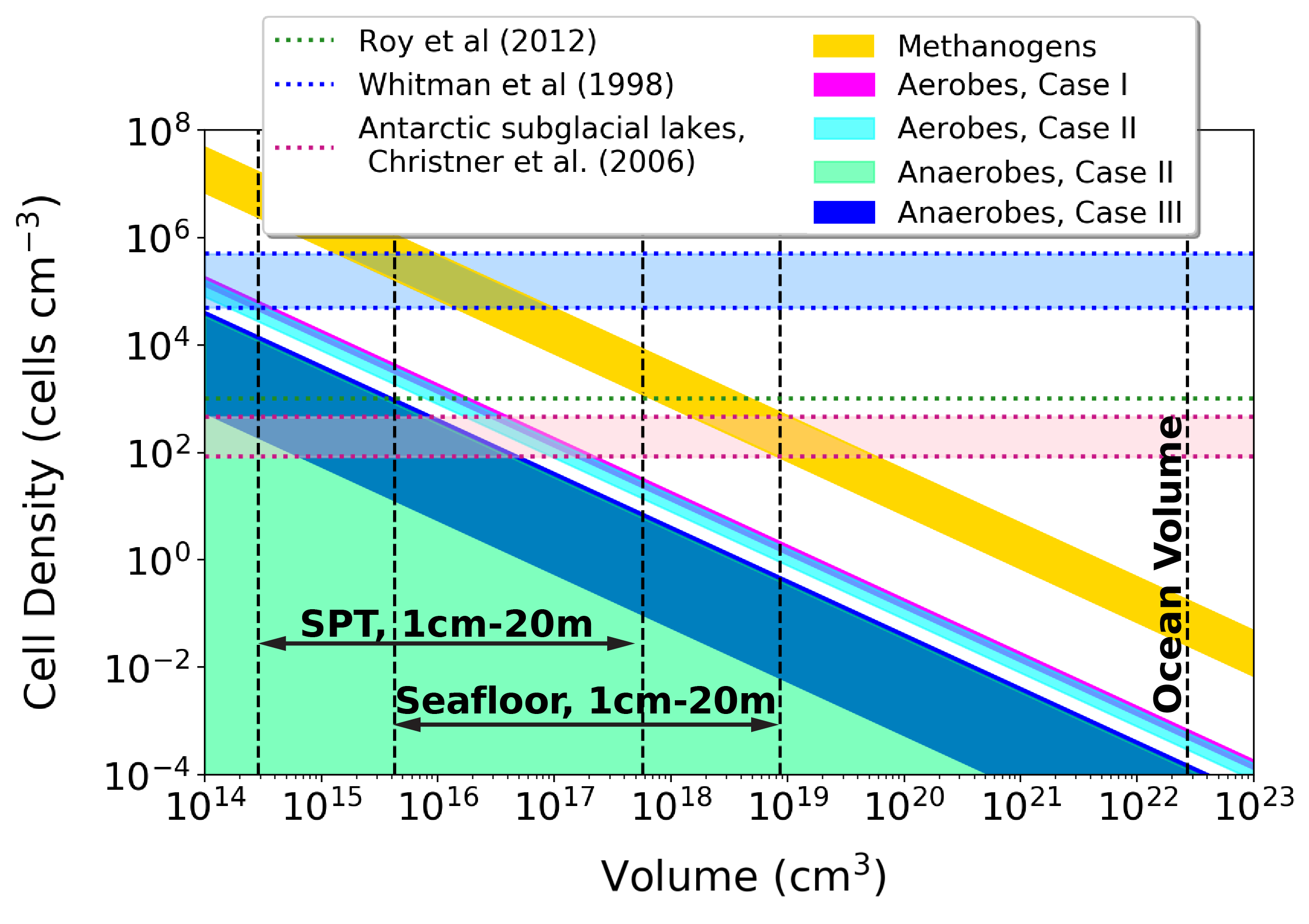}}
\caption{Hypothetical cell densities that could be sustained by each set of considered metabolisms on Enceladus. For reference, we have shown what cell densities might be present in the ocean if life were limited to only a specific region of the seafloor such as the area beneath the south polar terrain (SPT), the entire seafloor, or the entire ocean. For the former two regions, we have considered a range of depths in the seafloor from 1 cm to 20 m, for comparison to cell densities measured in Earth's shallow and deep oceans ($10^5$ and $10^4$ cells cm$^{-3}$, respectively, \cite{whitman1998prokaryotes}), in the Antarctica's Lake Vostok ($\sim$100-500 cells cm$^{-3}$, \cite{christner2006limnological}), and in the North Pacific Gyre 20 meters below the seafloor on Earth ($10^3$ cells cm$^{-3}$, \cite{roy2012aerobic}).}

\label{figone}
\end{figure}

\section{Conclusions}  
We have shown that the production of radiolytic oxidants on Enceladus could lead to redox disequilibria in the ocean, which could provide energy to support putative life. Radiolysis of surface ice, coupled with transport of ice to the ocean in the geologically active tiger stripe region, can deliver up to $9.4 \times 10^{15}$ moles of O$_2$ and $3.3 \times 10^{16}$ moles of H$_2$O$_2$. Electrons and gamma rays released as a result of the decay of $^{40}$K atoms in the ocean can produce another $4.2 \times 10^{16}$ moles of O$_2$ and $1.4 \times 10^{15}$ moles of H$_2$O$_2$ directly in the ocean. These radiolytic oxidants may build up in the ocean if the availability of reactive reductants is limited, or they could be converted to SO$_4^{2-}$ and ferric oxyhydroxides if water-rock interactions control the redox chemistry of the ocean. In the former case, the resulting ocean concentrations of oxidants (2.5 mmol O$_2$/kg H$_2$O or 3.1 mmol H$_2$O$_2$/kg H$_2$O) are much higher than the concentration range for CO$_2$ estimated in \cite{waite2017cassini} for pH 9 to 11, and thus these oxidants would likely have been detected by the INMS instrument on Cassini.

The absence of a non-ambiguous detection of O$_2$, along with existing evidence that Enceladus is hydrothermally active, instead support a scenario in which radiolytic O$_2$ and H$_2$O$_2$ are converted to SO$_4^{2-}$ and ferric oxyhydroxides, by oxidation of aqueous reductants in the ocean (Case II) and/or oxidation of reduced minerals at the seafloor (Case III). The end results of these two cases differ only in the relative abundances of SO$_4^{2-}$ versus ferric oxydroxides produced, and predict similar final SO$_4^{2-}$ concentrations ($\sim 1$mmol/kg H$_2$O or less, depending on ocean chemistry and seafloor mineralogy). Thus, the nominal 1 mmol/kg H$_2$O upper limit on [SO$_4^{2-}$] from the CDA instrument cannot be used to determine which process might dominate over the other. Ongoing laboratory experiments, coupled with continued analysis of the large dataset returned by Cassini CDA, are expected to further constrain the tentative upper limit for SO$_4^{2-}$ in Enceladus' ocean given here. An ambiguous detection of H$_2$S by INMS suggests that reactive reductants could be present, but whether these reductants can consume radiolytic oxidants before they are circulated through the rocky core depends on the relative time scales of each process. Further modeling of the hydrothermal flow of fluid through Enceladus' seafloor and deeper interior, along with future measurements of S isotopes, would help determine the extent to which each of these processes contributes to the reduction of radiolytic oxidants, and therefore how O$_2$ and H$_2$O$_2$ are partitioned into iron versus sulfide oxidation.

We have shown that, in addition to methanogenesis, aerobic and/or anaerobic reactions in each of our three cases can meet the minimum free energy requirement for terrestrial life, $\Delta G_{min}$, and provide maintenance energy to support cellular life within Enceladus. Although we have favored Cases II and III, aerobic metabolisms using oxidant fluxes close to those predicted in Case I could still support more than $\sim10^{19}$ cells at the ice-water interface where radiolytic oxidants from the surface are delivered. Aerobic metabolisms could also sustain another $\sim10^{19}$ cells in the ocean/seafloor if there are still low ($\sim10^{-19}-10^{-12}$ mol/kg H$_2$O in Case II) ocean concentrations of O$_2$ and/or H$_2$O$_2$. Anaerobic metabolisms could not sustain as much biomass as aerobic metabolisms due to their lower free energy yields, but at least some of these reactions are still viable and could support $10^{18}-10^{19}$ cells in Case II and Case III. Radiolytic oxidant production and redox chemistry in Enceladus' ocean and seafloor are therefore capable of supporting metabolic processes beyond methanogenesis, generating the possibility for a metabolically diverse microbial community in the ocean of Enceladus.

\clearpage

\appendix
\section{Oxidant Production on a Young Enceladus}
\begin{figure}[ht!]
\centerline{\includegraphics[height=2in]{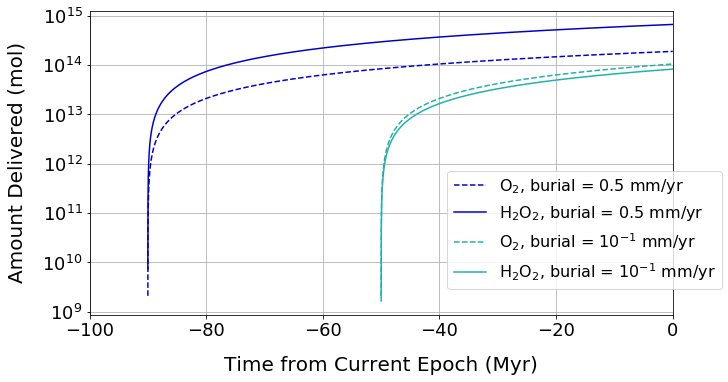}}
\caption{Cumulative delivery of O$_2$ (dashed lines) and H$_2$O$_2$ (solid lines) to Enceladus' ocean after 100 Myr for different average plume deposition rates. Curve offsets reflect increased delivery time through the ice shell with slower burial rates.}
\label{figone}
\end{figure}

\begin{figure}[ht!]
\centerline{\includegraphics[height=2.5in]{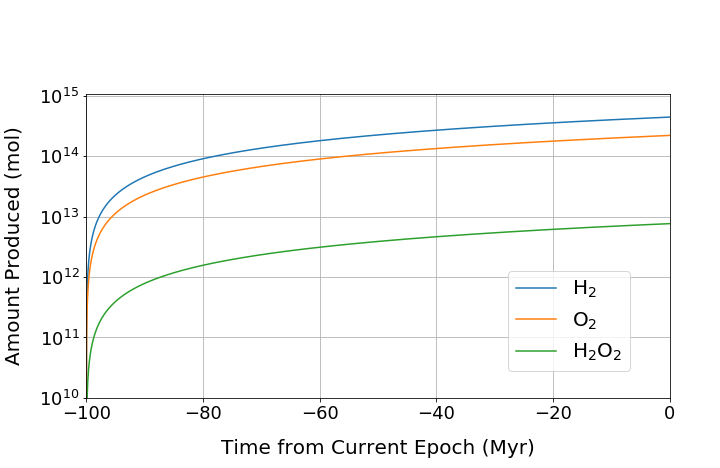}}
\caption{Cumulative moles H$_2$ (blue), O$_2$ (orange), and H$_2$O$_2$ (green) produced through radiation chemistry by the decay of $^{40}$K in Enceladus' ocean over 100 Myr.}
\label{figone}
\end{figure}
\clearpage

\begin{figure}[ht!]
\centerline{\includegraphics[height=2.5in]{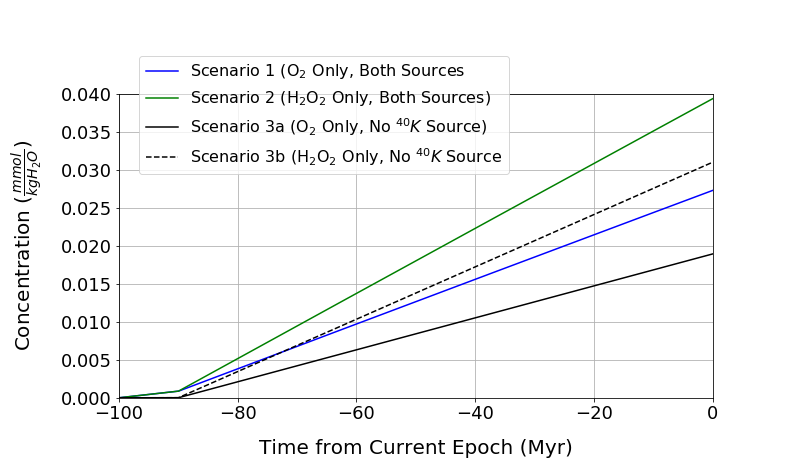}}
\caption{Concentration of O$_2$ or H$_2$O$_2$ in the ocean over 100 Myr, assuming it has not been depleted by reactions with reductants. Here, we show three scenarios for oxidant production: 1) radiolysis (in both the ice shell and in the ocean) forms only O$_2$ (solid blue line), 2) radiolysis forms only H$_2$O$_2$ (solid green line), or 3) no oxidants are formed by $^{40}$K decay as a result of back-reactions with H$_2$ into H$_2$O so that the ice shell is the only source of oxidants to the ocean, and yields either all O$_2$ (3a, solid black line) or all H$_2$O$_2$ (3b, dashed black line).}
\label{figone}
\end{figure}

\begin{figure}[ht!]
\centerline{\includegraphics[height=2in]{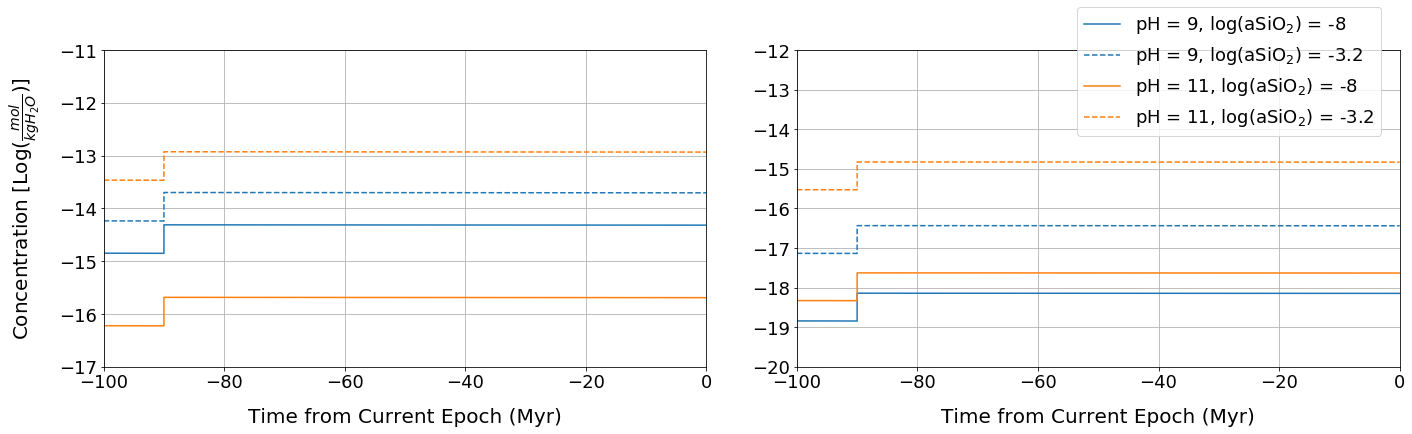}}
\caption{Steady-state concentration of O$_2$ (left panel) or H$_2$O$_2$ (right panel) in Case II over 100 Myr. pH 9 concentrations are shown in blue, and pH 11 concentrations are shown in orange. Solid lines correspond to log(aSiO$_2$ = -8), and dashed lines to log(aSiO$_2$ = -3.2).}
\label{figone}
\end{figure}

\begin{figure}[ht!]
\centerline{\includegraphics[height=3.5in]{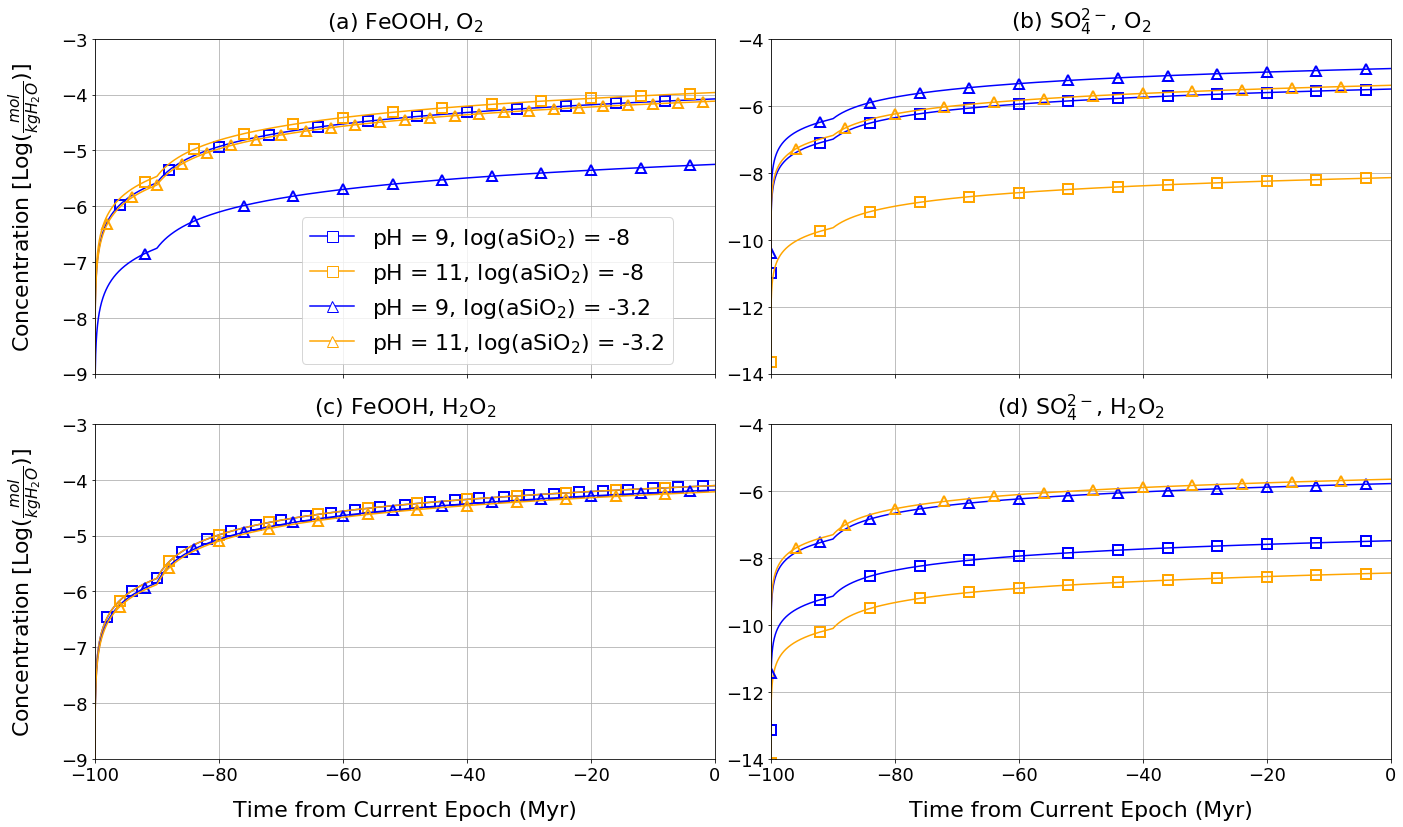}}
\caption{Concentration of FeOOH (a,c) and SO$_4^{2-}$ (b,d) produced by oxidation with O$_2$ (a,b) and H$_2$O$_2$ (c,d) in Case II over 100 Myr for pH 9 and log(aSiO$_2$) = -8 (blue squares), pH 11 and log(aSiO$_2$) = -8 (orange squares), pH 9 and log(aSiO$_2$) = -3.2 (blue triangles), and pH 11 and log(aSiO$_2$) = -3.2 (orange triangles). Note the y-axes scales are different for FeOOH than for SO$_4^{2-}$ in order to better see contrasts between different sets of conditions. For all considered conditions except oxidation by O$_2$ at pH 9 and log(aSiO$_2$) = -3.2, ferrous iron oxidation is strongly favored over sulfide oxidation.}
\label{figone}
\end{figure}

\begin{figure}[ht!]
\centerline{\includegraphics[height=1.7in]{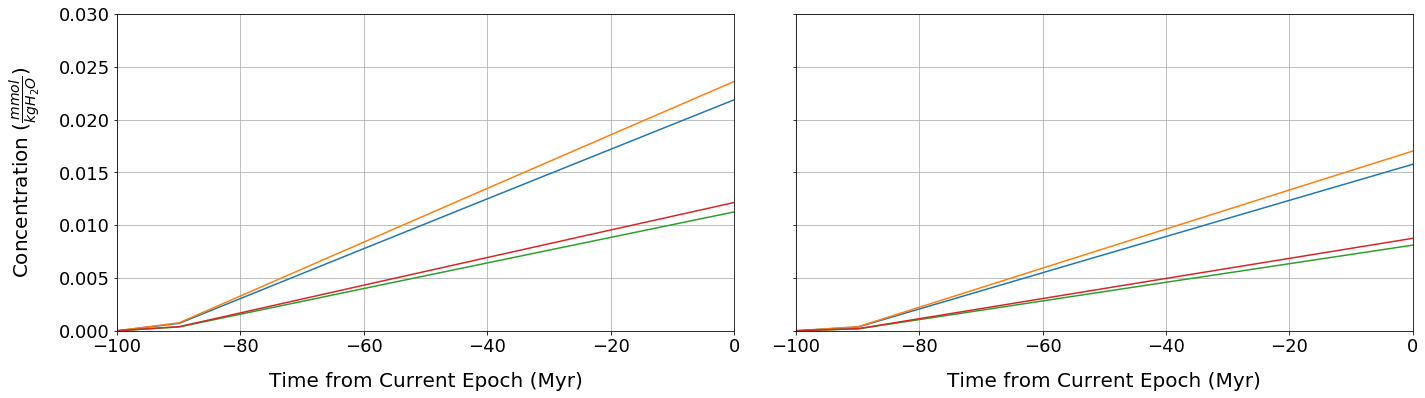}}
\caption{Concentration of FeOOH (blue for RHR, orange for OHR) and SO$_4^{2-}$ (green for RHR, red for OHR) produced by oxidation with O$_2$ (left panel) or H$_2$O$_2$ (right panel) in Case III over 100 Myr.}
\label{figone}
\end{figure}

\clearpage

\section{Oxidant Production Without a $^{40}K$ Source}

\begin{figure}[ht!]
\centerline{\includegraphics[height=2in]{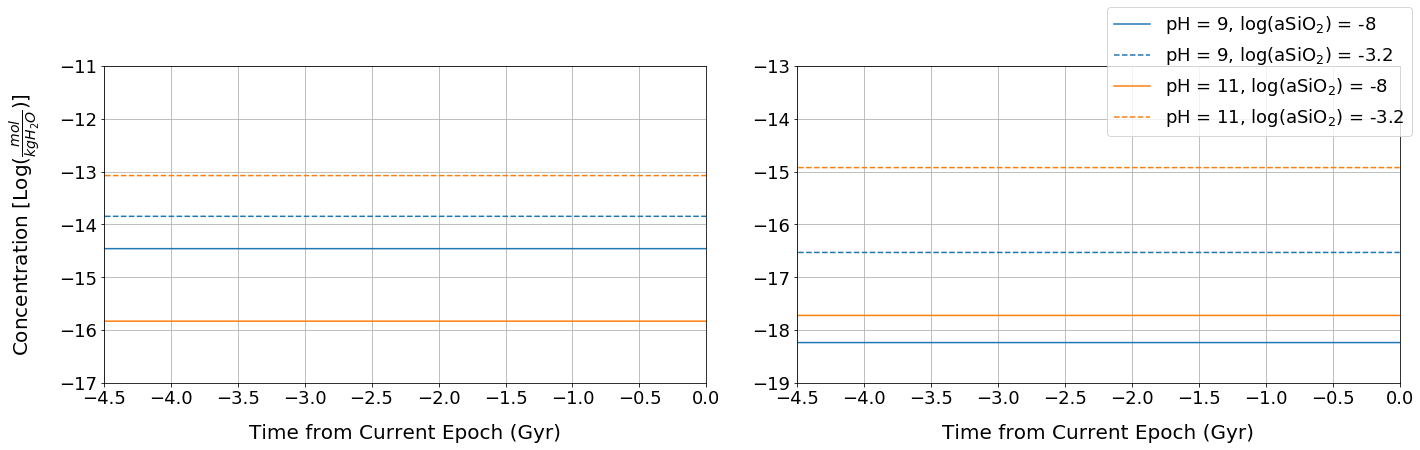}}
\caption{Steady-state concentration of O$_2$ (left panel) or H$_2$O$_2$ (right panel) in Case II over 4.5 Gyr if no oxidants are produced as a result of $^{40}K$ decay. pH 9 concentrations are shown in blue, and pH 11 concentrations are shown in orange. Solid lines correspond to log(aSiO$_2$ = -8), and dashed lines to log(aSiO$_2$ = -3.2).}
\label{figone}
\end{figure}

\begin{figure}[ht!]
\centerline{\includegraphics[height=3.3in]{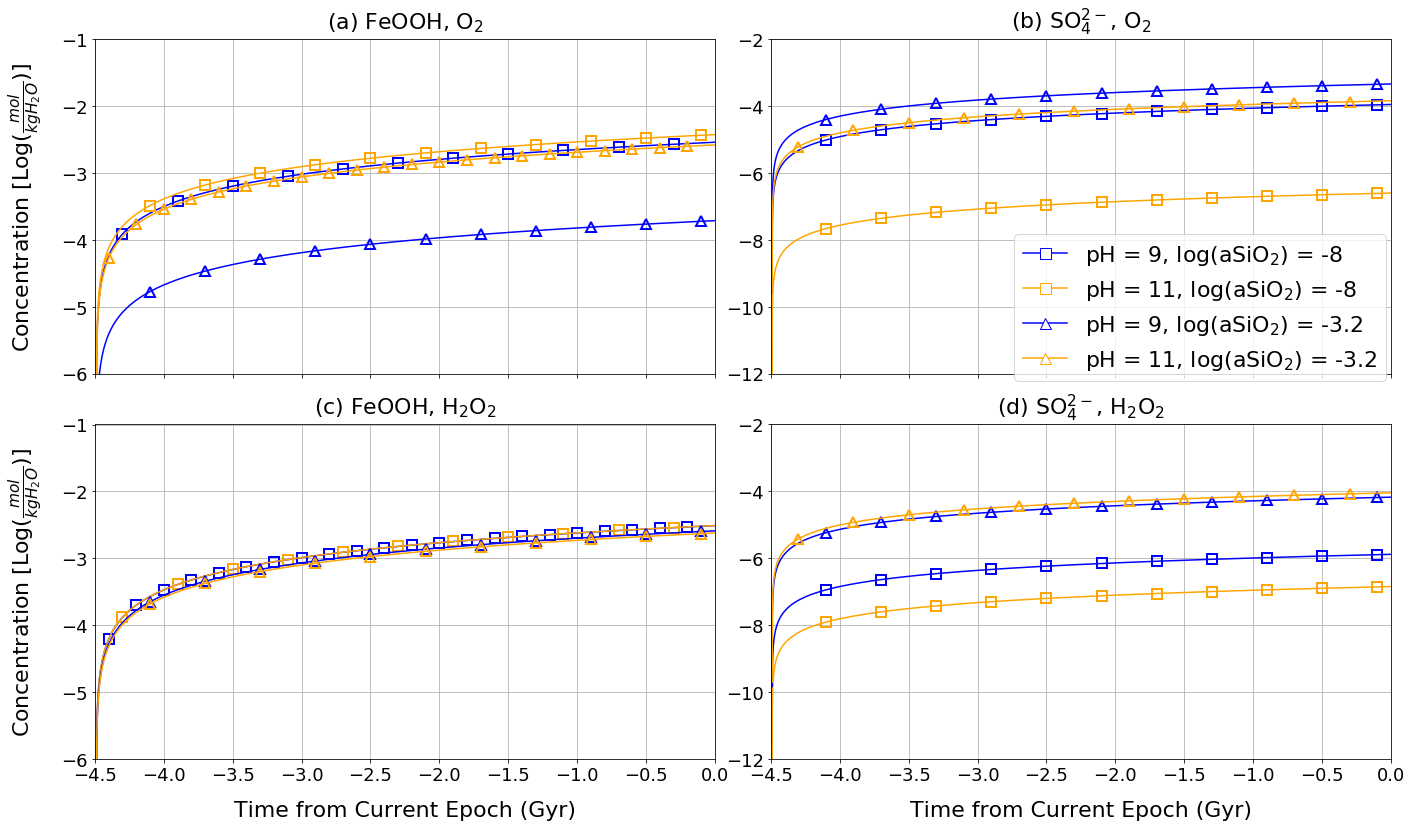}}
\caption{Concentration of FeOOH (a,c) and SO$_4^{2-}$ (b,d) produced by oxidation with O$_2$ (a,b) and H$_2$O$_2$ (c,d) in Case II over 4.5 Gyr for pH 9 and log(aSiO$_2$) = -8 (blue squares), pH 11 and log(aSiO$_2$) = -8 (orange squares), pH 9 and log(aSiO$_2$) = -3.2 (blue triangles), and pH 11 and log(aSiO$_2$) = -3.2 (orange triangles), if no oxidants are produced as a result of $^{40}K$ decay. Note the y-axes scales are different for FeOOH than for SO$_4^{2-}$ in order to better see contrasts between different sets of conditions. For all considered conditions except oxidation by O$_2$ at pH 9 and log(aSiO$_2$) = -3.2, ferrous iron oxidation is strongly favored over sulfide oxidation.}
\label{figone}
\end{figure}

\begin{figure}[ht!]
\centerline{\includegraphics[height=1.7in]{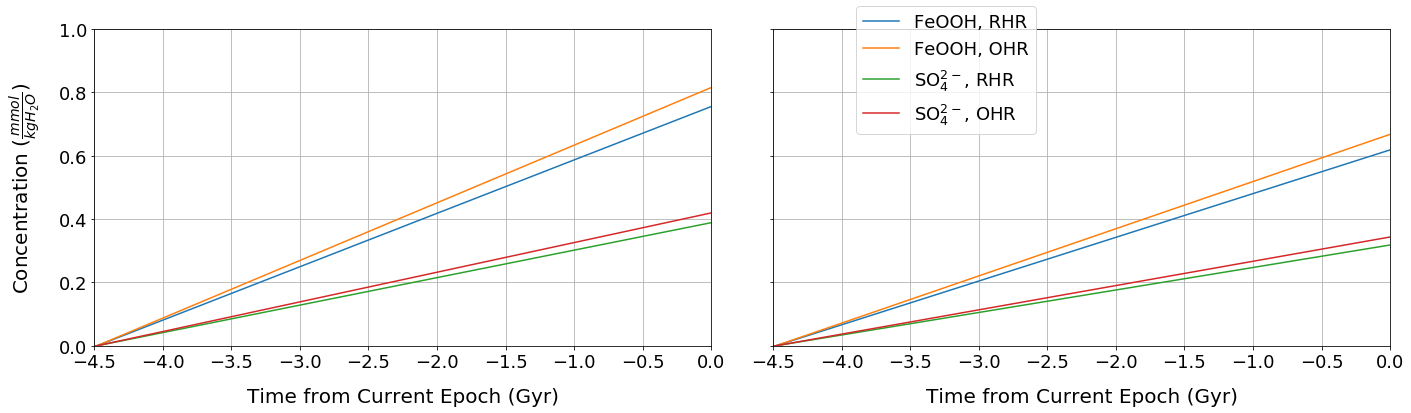}}
\caption{Concentration of FeOOH (blue for RHR, orange for OHR) and SO$_4^{2-}$ (green for RHR, red for OHR) produced by oxidation with O$_2$ (left panel) or H$_2$O$_2$ (right panel) in Case III over 4.5 Gyr, if no oxidants are produced as a result of $^{40}K$ decay.}
\label{figone}
\end{figure}

\clearpage

\section{Production Rates of Radiolytic Oxidants}
\begin{figure}[ht!]
\centerline{\includegraphics[height=2.5in]{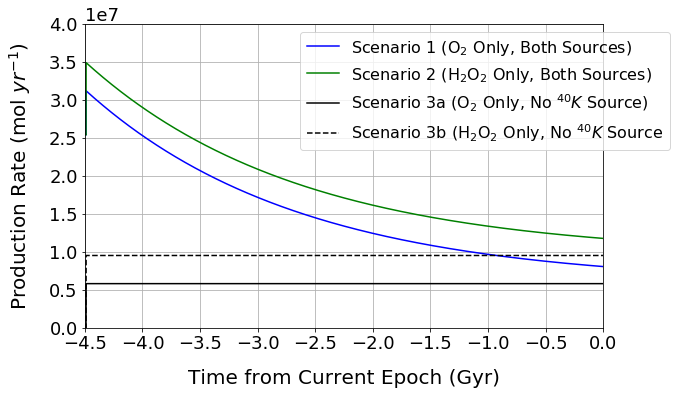}}
\caption{Production rates of O$_2$ and/or H$_2$O$_2$ over time.
 Here, we show three scenarios for oxidant production: 1) radiolysis (in both the ice shell and in the ocean) forms only O$_2$ (solid blue line), 2) radiolysis forms only H$_2$O$_2$ (solid green line), or 3) no oxidants are formed by $^{40}$K decay as a result of back-reactions with H$_2$ into H$_2$O so that the ice shell is the only source of oxidants to the ocean, and yields either all O$_2$ (3a, solid black line) or all H$_2$O$_2$ (3b, dashed black line).}
\label{figone}
\end{figure}

\section*{Acknowledgements}
This work was supported by the Cassini INMS subcontract from NASA JPL (NASA contract NAS703001TONMO711123, JPL subcontract 1405853) and the Cassini Project; C.R.G. was also supported by the NASA Astrobiology Institute. Support for T.M.H. was provided by NASA's Planetary Science Division Research Program. The work of F.P. was supported by the European Research Council (ERC Consolidator Grant 724908-Habitat OASIS). The NSF Center for Dark Energy Biosphere Investigations (C-DEBI OCE-0939564) and NASA ICEE2 grant 80NSSC19K0611 supported the participation of J.A.H.

\clearpage

\bibliography{Ray_Bibliography}

\end{document}